\documentclass[twocolumn,prx,superscriptaddress,longbibliography]{revtex4-1}
\usepackage{amssymb}
\usepackage{amsmath}
\usepackage{color}
\usepackage[colorlinks=true,urlcolor=blue,breaklinks,citecolor=blue]{hyperref}
\usepackage{graphicx}
\usepackage{braket}
\usepackage{subfigure}
\usepackage{inputenc}
\usepackage{xcolor}
\usepackage{dcolumn,bm}
\usepackage{cleveref}
\usepackage{float}
\usepackage{here}
\usepackage{hyperref}

\newcommand{\CSF}{C_\text{SF}}
\newcommand{\CDW}{C_\text{DW}}
\newcommand{\Cs}{C_\text{HI}}
\newcommand{\Cst}{C_\text{HI}}
\newcommand{\chimax}{\chi_\text{max}}

\usepackage[draft]{changes}
\usepackage{ulem} 

\begin{document}
\title{Supersolid-Superfluid phase separation in the extended Bose-Hubbard model}

\author{Korbinian Kottmann}
\affiliation{ICFO - Institut de Ciencies Fotoniques, The Barcelona Institute of Science and Technology, Av. Carl Friedrich Gauss 3, 08860 Castelldefels (Barcelona), Spain}
\author{Andreas Haller}
\affiliation{Department of Physics and Materials Science, University of Luxembourg, 1511 Luxembourg, Luxembourg}
\affiliation{Institute of Physics, Johannes Gutenberg University, D-55099 Mainz, Germany}
\author{Antonio Ac\'in}
\affiliation{ICFO - Institut de Ciencies Fotoniques, The Barcelona Institute of Science and Technology, Av. Carl Friedrich Gauss 3, 08860 Castelldefels (Barcelona), Spain}
\affiliation{ICREA, Pg. Lluís Companys 23, 08010 Barcelona, Spain}
\author{Grigory E. Astrakharchik}
\affiliation{Departament de F\'isica, Campus Nord B4-B5, Universitat Polit\`ecnica de Catalunya, E-08034 Barcelona, Spain}
\author{Maciej Lewenstein}
\affiliation{ICFO - Institut de Ciencies Fotoniques, The Barcelona Institute of Science and Technology, Av. Carl Friedrich Gauss 3, 08860 Castelldefels (Barcelona), Spain}
\affiliation{ICREA, Pg. Lluís Companys 23, 08010 Barcelona, Spain}

\begin{abstract}
Recent studies have suggested a new phase in the extended Bose-Hubbard model in one dimension at integer filling \cite{Kottmann2020, Batrouni}. 
In this work, we show that this new phase is phase-separated into a supersolid and superfluid part, generated by mechanical instability.
Numerical simulations are performed by means of the density matrix renormalization group algorithm in terms of matrix product states. 
In the phase-separated phase and the adjacent homogeneous superfluid and supersolid phases, we find peculiar spatial patterns in the entanglement spectrum and string-order correlation functions and show that they survive in the thermodynamic limit.
In particular, we demonstrate that the elementary excitations of the homogeneous superfluid with enhanced periodic modulations are phonons, find the central charge to be $c=1$, and show that the velocity of sound, extracted from the intrinsic level splitting for finite systems, matches with the propagation velocity of local excitations in dynamical simulations.
This suggests that the low-energy spectrum of the phase under investigation is effectively captured by a spinless Luttinger liquid, for which we find consistent results between the Luttinger parameter obtained from the linear dependence of the structure factor and the algebraic decay of the one-body density matrix.
\end{abstract}

\maketitle

\section{Introduction}

\noindent{\it Motivation.} Bosonic Hubbard models remain in the focus of interest in condensed matter and ultracold quantum matter physics since the seminal paper of Fisher {\it et al.}~\cite{Fisher}. In recent years, considerable attention was devoted to extended/non-standard Hubbard models (for a review cf.~\cite{Gajda}). There is a number of reasons for this:
\begin{itemize}
\item{\bf Fundamental interest.} Extended Bose Hubbard models provide perhaps the simplest models that include beyond on-site interactions.
\item{\bf Richness of quantum phases.} They exhibit a plethora of quantum phases arising due to the interactions, even in one dimension (1D): Mott insulator (MI), Haldane insulator (HI), superfluid (SF), supersolid (SS), and charge density wave (CDW).
\item{\bf Long-range interactions.} They provide the first step towards a description of systems with long-range interactions, such as dipolar ones, for instance.
\item{\bf Experimental feasibility.} Quantum simulators of these models and their variants are experimentally feasible in various platforms: ultracold atoms/molecules in optical lattices \cite{lewenstein2012ultracold}, systems of trapped ions, Rydberg atoms, etc.
\end{itemize}

\noindent{\it State of art.} This work deals with the physics of the extended Bose Hubbard model in 1D and focuses on three of the most challenging and discussed phenomena of contemporary physics: supersolidity, phase separation, and entanglement. Supersolidity in the extended Hubbard model in 1D has been studied previously for incommensurate fillings~\cite{Kawaki2017,Kuehner1997,Kuehner1999,Mishra2009}, and was claimed to be found for filling 1 in Ref.~\cite{Deng2011} without in-depth discussion, however. The complete phase diagram of the model was described by Batrouni {\it et al.} (see~\cite{Batrouni} and references therein; our work expands the results of Ref.~\cite{Rossini2012}). These authors studied the phase diagram of the one-dimensional bosonic Hubbard model with contact ($U$) and nearest-neighbor ($V$) interactions focusing on the gapped HI phase which is characterized by an exotic nonlocal order parameter. They used the Stochastic Green Function quantum Monte Carlo as well as the Density Matrix Renormalization Group (DMRG) algorithm to map out the phase diagram. Their main conclusions concern the existence of the HI at filling factor $\nu=1$, while the SS phase exists for a very wide range of parameters (including commensurate fillings) and displays power-law decay in the one-body Green function. In addition, they found that at fixed integer density, the system exhibits phase separation in the $(U,V)$ plane.\\

\noindent{\it Our results.}
In this work we apply the state-of-art DMRG method in terms of Tensor Networks, i.e. Matrix Product States (MPS) to study the ground-state properties of the extended Bose-Hubbard model,
\begin{multline}
\label{eq:H}
H = -t \sum_i \left( b^\dagger_i b_{i+1} + b^\dagger_{i+1} b_i \right) \\
+ \frac{U}{2} \sum_i n_i(n_i -1) + V \sum_i n_i n_{i+1},
\end{multline}
with nearest neighbour interaction on a one dimensional chain with $L$ sites. Here, $n_i = b^\dagger_i b_i$ is the number operator for Bosons defined by $[b_i,b_j^\dagger] = \delta_{ij}$. The model is characterized by three energy scales: the nearest-neighbor tunneling amplitude $t$, on-site interactions of strength $U$, and nearest-neighbor interactions tuned by $V$.
We set the energy scales in units of the tunneling coefficient by setting $t=1$ and continue with dimensionless quantities.

We perform simulations both in a microcanonical ensemble (fixed number $N$ of particles) and a canonical ensemble (fixed chemical potential $\mu$ and fluctuating number of particles).

We expand the results of Ref.~\cite{Batrouni,Kottmann2020} in several aspects, which can be summarized as follows:
\begin{itemize}
\item{\it Phase separation (PS)}. 
For weak on-site interactions (small $U$) and strong nearest-neighbor ones (large $V$), the ground state for filling $n$ close to unity corresponds to a phase separation between SF and SS phases. 
The characteristics of the phase-separated ground state are illustrated in fig.~\ref{fig:SS_SF_separation}.
We perform a numerical analysis of the mechanical stability in terms of second derivatives of the energy and the Gibbs potential as a function of the density $n$. 
The nature of the SF-PS and PS-SS transitions is discussed in fig.~\ref{d2E-S_f}.

\item{\it Phase coexistence}. The phase-separated ground state corresponds to a genuine phase co-existence, stable for quite a relevant interval of values of the mean density $n$. We also study $n$ as a function of the chemical potential  $\mu$ for fixed $U$ ($V$) and varying $V$ ($U$) in fig.~\ref{fig:n-mu-U}.

\item{\it Entanglement oscillations.} In the SF phase, in the regime of parameters corresponding to the phase separation/phase coexistence, all single-particle observables seem to be spatially homogeneous, while entanglement R´enyi entropies and entanglement spectra exhibit oscillations. The spatial period of these oscillations, as well as the period of the Schmidt gap closing, is of the order of 10-20 lattice constants, i.e. has nothing to do with the periodicity of the CDW or SS, which is 2 lattice constants.

\item{\it Luttinger liquid picture.} 
Both in SF and SS phases, the excitation spectrum is governed by gapless linear phonons which make the Luttinger liquid description applicable and therefore allow predictions of the long-range behavior of the correlation functions.

We use Luttinger liquid theory to explain the presence of oscillations in the entanglement spectrum thus ruling out the possibility that these oscillations appear due to topological effects.
\end{itemize}

\noindent{\bf Plan of the paper}. In Sec.~\ref{sec:prelim} we discuss shortly the aspects of the present study from a general perspective: supersolidity, phase separation, topology, and entanglement properties in many-body systems in 1D.
In~Sec.~\ref{sec:simulation} we describe our numerical methods and in Sec.~\ref{sec:hamiltonian} recap the phenomenology of the system. Sections~\ref{sec:phase-separation},\ref{sec:oscillations} are devoted to the detailed discussion of our results concerning phase separation, including the analysis of the mechanical stability, and entanglement oscillations. Luttinger theory is discussed in Sec.~\ref{sec:luttinger}, while we conclude shortly in Sec.~\ref{sec:conclusions}.

\section{Preliminaries: Phenomena of interest}
\label{sec:prelim}
\noindent \textbf{Supersolidity.} A supersolid is a spatially ordered material with superfluid properties. Practically, since the discovery of superfluidity by Kapitza, Allen, and Misener~\cite{Kapitza1938,Allen1938}, there have been constant efforts to predict and realize systems that exhibit supersolidity. 
In 2004 an observation of a finite superfluid signal in solid helium was reported in Refs.~\cite{KimScience2004,KimNature2004}, while this claim eventually was disproved, it attracted an additional interest to supersolids and eventually, their formation has been observed in other systems.
Several mechanisms and scenarios for supersolidity were proposed from superfluid Helium to ultracold atoms:
\begin{itemize}
\item \textit{Andreev-Lifshitz-Chester scenario} \cite{Andreev69,Chester70}. 
In this scenario, vacancies (i.e. empty sites normally occupied by particles in a perfect crystal) exist even at absolute zero temperature. 
These vacancies might be caused by quantum fluctuations, which also cause them to move from site to site. 
Because vacancies are bosons, if such clouds can exist at a very low temperature $T$, then a Bose-Einstein condensation of vacancies could occur at temperatures less than a few tenths
of a kelvin.
\item {\it Shevchenko scenario} \cite{Shevchenko87,Boninsegni2007,Pollet2008}.  Here, mass flow occurs along dislocation cores forming a three-dimensional (3D) network.
\item {\it Supersolid stripe phase} \cite{Li2017}. This phase can be formed in dilute weakly interacting two-component Bose gases with spin-orbit coupling.
\item {\it Supersolid in a cavity} \cite{Leonard17}. One can realize a  supersolid with the breaking of continuous translational symmetry that emerges from two discrete spatial ones by symmetrically coupling a Bose-Einstein Condensate to the modes of two optical cavities.
\item {\it Metastable supersolid phase} \cite{Tanzi2019,Bottcher2019,Chomaz2019} Such phase was observed in systems of dipolar quantum droplets.
\item{\it Lattice supersolids} \cite{Kuehner1997,Kuehner1999,Mishra2009,Deng2011,Kawaki2017}. This mechanism occurs in systems described by extended Hubbard models; it is particularly efficient in the Hubbard models with long-range interactions, such as, for instance, dipolar interactions~\cite{Goral2002,Capogrosso2010,Hauke2010,Maik2012}. Due to the next-to nearest neighbor or even longer range repulsion, atoms tend to crystallize occupying a commensurate fraction of sites (with one or few atoms in an occupied site). Such states are termed density wave states (DW), and they are fully analogs of Mott insulator states (MI) with fully localized atoms, but at lower densities. Quantum fluctuations may melt these crystals, leading to the formation of supersolids. Such a kind of superpersolidity in the extended Hubbard model in 1D has been studied previously for incommensurate fillings~\cite{Kawaki2017,Kuehner1997,Kuehner1999,Mishra2009}, and was claimed to be found for filling 1 in~\cite{Deng2011} without further discussion.
\end{itemize}

\noindent{\bf Phase separation.} Phase separation is at the center of interest of physics for decades. It is understood as the creation of two distinct phases from a miscible homogeneous mixture.
A paradigmatic example of phase separation is between two immiscible liquids such as oil and water. Recently, two kinds of phase separation instances became very hot subjects in science: {\it liquid-liquid phase separation} in biology as a regulator of cellular biochemistry~(\cite{Banani2017}, see also~\cite{Munoz2020} and references therein), and
{\it quantum phase separation}. Classically, these processes occur typically via two distinct mechanisms:
\begin{itemize}
\item{\it Spinodal decomposition} \cite{Binder1987}. Spinodal decomposition takes place when the decomposition into two phases occurs with no nucleation barrier. The mixture is initially in an unstable state so that fluctuations in the system spontaneously grow to reduce the free energy. In the quantum scenario, the decay of unstable states may lead to the macroscopic amplification of those quantum fluctuations that initiated the process (cf.~\cite{Haake1978,Glauber1978}).

\item{\it Nucleation}. In nucleation and in the associated growth, there is a nucleation barrier. While in spinoidal decomposition an unstable phase corresponds to the maximum of the free energy, nucleation and growth occurs in a metastable phase and is resistant to small fluctuations. 
\end{itemize}

In quantum mechanics, phase separation typically concerns conducting (metallic, superfluid) and insulating phases. A characteristic example is the formation of the ``wedding cake'' structures in a system described by the  Bose-Hubbard (BH) model in an optical lattice in a loose harmonic trap~\cite{Kato2009,lewenstein2012ultracold}. In such a case, MI regions with a fixed number of atoms per lattice site are separated by SF rings (``wedding cake'' structure). Locally, the state of the system is determined by the trapping potential, which acts as local chemical potential. Note, however,  that for a fixed number of atoms in the  homogeneous system, the ground state of the BH model is always SF if the number of atoms $N$ is incommensurable with the number of lattice sites. In  a strict sense, this is not a genuine phase separation, since it does not lead to phase coexistence in a spatially homogenous system. This will be different in the extended BH model studied in this work.  \\
work. 

It is worth mentioning that similar mechanisms lead to "wedding cake" structures in extended Bose Hubbard models \cite{Dutta15}, such as dipolar Hubbard models \cite{Lahaye09,Capogrosso10}. Another relevant example might be the itinerant ferromagnetic instability in repulsive fermions, studied primarily in trapped gases \cite{LeBlanc09,Pilati10,Penna17,Amico18}, but investigated also very intensively in extended (and in particular dipolar) Fermi Hubbard models, where “wedding cake” structures are also expected to appear  \cite{Cannon90,Si93,Onari04,Chang10,Wu14,Szalowski17,Sompet21}.


\noindent{\bf Topology in 1D}. Since we are going to argue that the considered model does not possess topological order in SS and SF phases, let us remind the reader about the peculiarity of low-dimensional systems. 
In 1D, topological order exists only in the form of symmetry-protected topological order (SPTP).
There are various ways of characterizing topological order: it is common to look at topological invariants, edge states, hidden order parameters, and entanglement properties, i.e. entanglement entropies~\cite{Eisert2010} and entanglement spectrum (ES)~\cite{Li2008}. There are two paradigmatic models that exhibit topological order in 1D: the Su-Schrieffer-Heeger (SSH) model~\cite{Su1979,Su1980} and related models such as the original model of the acetylene chain with electrons interacting with phonons living on the bonds, or families of bosonic models in dynamical lattices, where spins on the bond mimic phonons~ \cite{Gonzalez2018,Gonzalez2019}), and the Affleck-Kennedy-Lieb-Tasaki  model \cite{Affleck1987} (or related models such a biquadratic-bilinear Heisenberg model, cf. \cite{DeChiara2012,Lepori2013} and references therein). 
The latter phase is often termed as Haldane phase, since it was postulated and discovered by Haldane, in the context of $S=1$ 1D Heisenberg model \cite{Haldane83a,Haldane83b}. Haldane formulated there the famous Haldane \textit{conjecture}, saying the 1D Heisenberg  models for half-integer spins are gapless with algebraically decaying correlations, whereas those with integer spins are gapped with exponentially decaying correlations. The AKLT model, which is an example of biquadratic-bilinear Heisenberg models, has the same properties. Its ground state in the case of open boundary conditions is four-fold degenerate due to the hidden $D_2=Z_2\times Z_2$ symmetry \cite{DerNijs89}. Kennedy and Tasaki \cite{K&T92} demonstrated this by applying a non-local unitary to the Hamiltonian of the models, as one gets a model with explicit global discrete symmetry ($\Pi$-rotation about $x,y,z$ axes $=D_2= Z_2\times Z_2$). Ferromagnetic order in the transformed system corresponds to non-local topological order in the original Heisenberg model. This order is indeed protected by the $D_2$ symmetry \cite{Pollmann2010c}. The model possesses also two more discrete symmetries: time reversal and space inversion about the bond centre. Adding various terms to the Hamiltonian, one may break some of these, but as long as one of them survives, so does the Haldane phase. 

Later it was shown that if we allow arbitrary deformations, there exists only one phase in 1D; the reasonable classification of quantum phases can be achieved only applying deformation that respect discrete symmetries \cite{wen1}; in fact, full classification of symmetry proitected quantum phases, including spin, fermionic and bosonic systems was achieved \cite{chen11,chen11a,turner11,Fidkosvki11,schuch11}.

The SSH-like and AKLT-like models several quantum
phase transitions, and have the following properties with
respect to topological order:
\begin{itemize}
\item{\it Topological invariants}. The winding number characterizes very well the topological phases of the SSH-family (cf.~\cite{Maffei2018}). These topological invariants can be, but are more rarely, used for the AKLT-family.
\item{\it Edge states}. The bulk-edge correspondence works obviously very well for the SSH-family. For the AKLT-family it requires a numerical solution with open boundary conditions but is also straightforward.
\item{\it Hidden order parameters}. A string order parameter is typically defined and used for the AKLT-family. It exhibits long range order in the topological phase, and decays exponentially or vanishes in other, non-topological phases.
\end{itemize}

\noindent{\bf Entanglement properties}. 
Here we summarize the properties of entanglement in many-body systems in 1D, analyzed later in the manuscript. 
We pay special attention to possible sources of spatial oscillations of entanglement entropies and/or spectrum.
\begin{itemize}
\item{\it Entanglement entropies/spectrum far from criticality}. In conventional uniform systems, these quantities are homogeneous. Obviously, in systems that are ``dimerized'' (trimerized, quadrumerized, etc.), entanglement entropies/spectrum oscillate, even though all single-particle observables are spatially homogeneous. In the extreme case, $k$-merized states are $k$-producible: they are defined by the product of entangled states of size $k$. If we put the cut between the $k$-mers, we get entropies equal to zero, and a trivial entanglement spectrum corresponding to a product state. Interestingly, their results also hold for disordered systems, as shown recently in Ref.~\cite{Tan2020}, using strong disorder renormalization group methods.

  \item{\it  Entanglement entropies at criticality}. For standard systems in 1D of finite (but large) size $L$ and open boundaries, the entanglement entropies of ground states read
  \begin{equation}
    S_\alpha(\ell)= -\ln({\rm Tr}\rho^\alpha(\ell))/(1-\alpha)
    \label{eq:renyi_entropy}
  \end{equation}
  
  where $\rho^\alpha(\ell)$ is the reduced density matrix of the block of size $\ell$ \cite{Calabrese09}.
      The behaviors of R´enyi entropies for ground states of critical (gapless) systems are well known according to conformal field theory~\cite{calabrese2004,calabrese2009}
\begin{eqnarray}
S_\alpha(\ell) &=& \frac{\alpha+1}\alpha\frac c{6b} \ln\left( d[\ell|L] \right) + S_{\rm sl} + \gamma, \label{Renyi_scaling}\\
d[\ell|L] &=& \left|L/\pi\sin\left( \pi \ell/L \right)\right|,
\end{eqnarray}
where d$[\ell|L]$ is the chord length on a ring of perimeter $L$. The leading part exhibits a universal scaling law with prefactor factor $c$ called the central charge of the conformal field theory (in fermionic systems, it is equal to the number of Fermi points).
An additional factor $b$ distinguishes the case of periodic ($b=1$) and open ($b=2$) boundary conditions and $\gamma$ constitutes a non-universal constant. Sub-leading terms are denoted by $S_{\rm sl}$ and, in general, oscillate in space.

    \item{\it  Entanglement spectrum at criticality}. The Schmidt gap (difference between the lowest and  the second-lowest eigenvalue of the ES, or between the highest two squared Schmidt coefficients) closes, i.e. ceases to zero. In topological phases, the ES remains degenerated, in accordance with the symmetry protecting the topological order -- it was first demonstrated for the AKTL-family in Ref. \cite{Pollmann2010}. In the case of oscillating ES, as in the SSH-family, the oscillations cease to zero at criticality \cite{Tan2020}. Finally, if we approach criticality from a trivial phase, where there exists a ``standard'' local order parameter (magnetization, staggered magnetization, etc.), then closing of the Schmidt gap is directly related to the vanishing of the order parameter at criticality \cite{DeChiara2012,Lepori2013}.
 \end{itemize}

\section{Simulation Method}
\label{sec:simulation}
We calculate the ground states by means of the DMRG algorithm expressed in terms of MPS states~\cite{SCHOLLWOCK2011,Orus2013}. A general multipartite state of $L$ parties, with (finite) local dimension $d$, $\ket{\Psi} = \sum_{\bm{\sigma}} c_{\bm{\sigma}} \ket{\bm{\sigma}}$, where $\bm{\sigma}=\sigma_1\ldots\sigma_ L$ is the vector of local indices $\sigma_i = 1,\ldots,d$, can always be decomposed into products of tensors with the aid of the singular value decomposition. We use the convention of Vidal \cite{Vidal2003} for a canonical form, and write our ground state in the MPS form

\begin{multline}
\label{eq:ansatz_mps}
\ket{\Psi} = \sum_{\bm{\sigma}} \Gamma^{\sigma_1} \bm{\lambda}^{[1]} \cdots \bm{\lambda}^{[i-1]} \Gamma^{\sigma_i} \bm{\lambda}^{[i]} \cdots  \\
\bm{\lambda}^{[L-1]} \Gamma^{\sigma_{L}} \ket{\sigma_1 \ldots \sigma_i \ldots \sigma_{L}}.
\end{multline}

At site $i$, $\{\Gamma^{\sigma_i}\}$ is a set of $d$ matrices and $\bm{\lambda}^{[i]} = \text{diag}(\lambda_1,\lambda_2,\cdots,\lambda_{\chimax})$  the diagonal singular value matrix of a bipartition of the chain between site $i$ and $i+1$, i.e. the Schmidt values (see \cite{SCHOLLWOCK2011}). One then approximates the exact ground state by keeping only the $\chimax$ largest Schmidt values for each partition, where $\chimax$ is known as the bond dimension. This is the best approximation of the full state in terms of the Frobenius norm and enables us to handle big system sizes. Note that when simulating Bosonic systems that are in principle infinite dimensional, we introduce an error by truncating the local Hilbert space dimension to $d$. One has to be careful in choosing this parameter and make sure that the physics is still captured within the chosen truncation, e.g. by scaling in this parameter and comparing the results as is shown in the supplementary materials (SM) \cite{suppl}.
Eq. (\ref{eq:ansatz_mps}) corresponds to finite length and open boundary conditions. The DMRG algorithm can also be formulated in the thermodynamic limit for infinite MPS (iMPS) \cite{Vidal2006,McCulloch2008,Orus2008}. In this case, instead of a finite chain, we have a finite and repeating unit cell of length $L_\infty$.
For calculations in the microcanonical ensemble with a fixed number of particles, we can explicitly target the ground state for a filling $n :=\sum_i \braket{n_i}/L$ by employing $U(1)$ symmetric tensors \cite{Silvi2017}, which is implemented in the open source library \textsf{TeNPy} \cite{tenpy}. 

\section{Observables}
\label{sec:hamiltonian}
The extended Bose-Hubbard model, \cref{eq:H}, admits a rich phase diagram and has been investigated thoroughly in the past two decades \cite{Rossini2012, Deng2013, Kuehner1997, Kuehner1999, Mishra2009, Urba2006, Ejima2014, Cazalilla2011, Batrouni2006, Deng2011, Berg2008}.

In these works, the following expectation values were analyzed in order to classify the observed phases:
\begin{align}
\CSF(i,j) &= \braket{b^\dagger_i b_j} \label{eq:CSF} \\
\CDW(i,j) &= \braket{\delta n_i (-1)^{|i-j|} \delta n_j} \label{eq:CDW} \\
\Cs(i,j) &= \braket{\delta n_i \exp\left( -i \pi \sum_{i \leq l \leq j-1} \delta n_l\right) \delta n_j}
\label{eq:Cs}
\end{align}
for which $\delta n_i = n_i - n$. 
In one spatial dimension, the observable $\CSF$ discriminates between the Mott-insulating (MI) phase and the superfluid (SF) phase by means of an exponential resp. power-law decay. A power-law decay of $\CSF$ is considered long-range correlations in the special case of one spatial dimension. For two or more spatial dimensions, a superfluid is characterized by true long-range correlations in $\CSF$.
The other two functions, $\CDW$ and $\Cs$, show true long-range correlations and assume constants at long distances $1\ll |i-j|$ in their corresponding phases, i.e. $\Cs$ signals the HI while $\CDW$ -- CDW ordering \cite{Rossini2012}. Note that $\Cs$ vanishes in Mott-like phases, where $\delta n_i=0$. It decays exponentially in the SF and SS phases - there $\delta n_i$ can be regarded as independent random variables so that the average of the exponential factor in $\Cs$ must decay as $\eta^{|i-j|}$ with $|\eta|\le 1$. If $\delta n_i$ are correlated, as in Luttinger liquids in 1D, the decay also has a power law and $\CDW$ CDW ordering~\cite{Rossini2012}. 
In the present paper we concentrate on SF, SS, and the phenomenon of phase separation, but the considered model exhibits also the HI phase, in which $\Cs$ shows long range order, in full agreement to Ref. \cite{Batrouni}. We do, however, evaluate $\Cs$ in the SF and SS phases, to demonstrate that there $\Cs$ decays with a power law and reveals spatial features from the entanglement spectrum (yet no topological order is determined). 


Recently, machine learning has been used to detect the presence of new phases in the region of strong nearest-neighbor and weak on-site interactions where the system phase-separates into a superfluid and a supersolid phase~\cite{Kottmann2020}. 

In the following, we explore and provide a detailed description of this region of the phase diagram considering in detail different fillings, bringing attention to interesting features of the entanglement distribution.

\section{Phase Separation}
\label{sec:phase-separation}
\begin{figure}
\centering
\includegraphics[width=.48\textwidth]{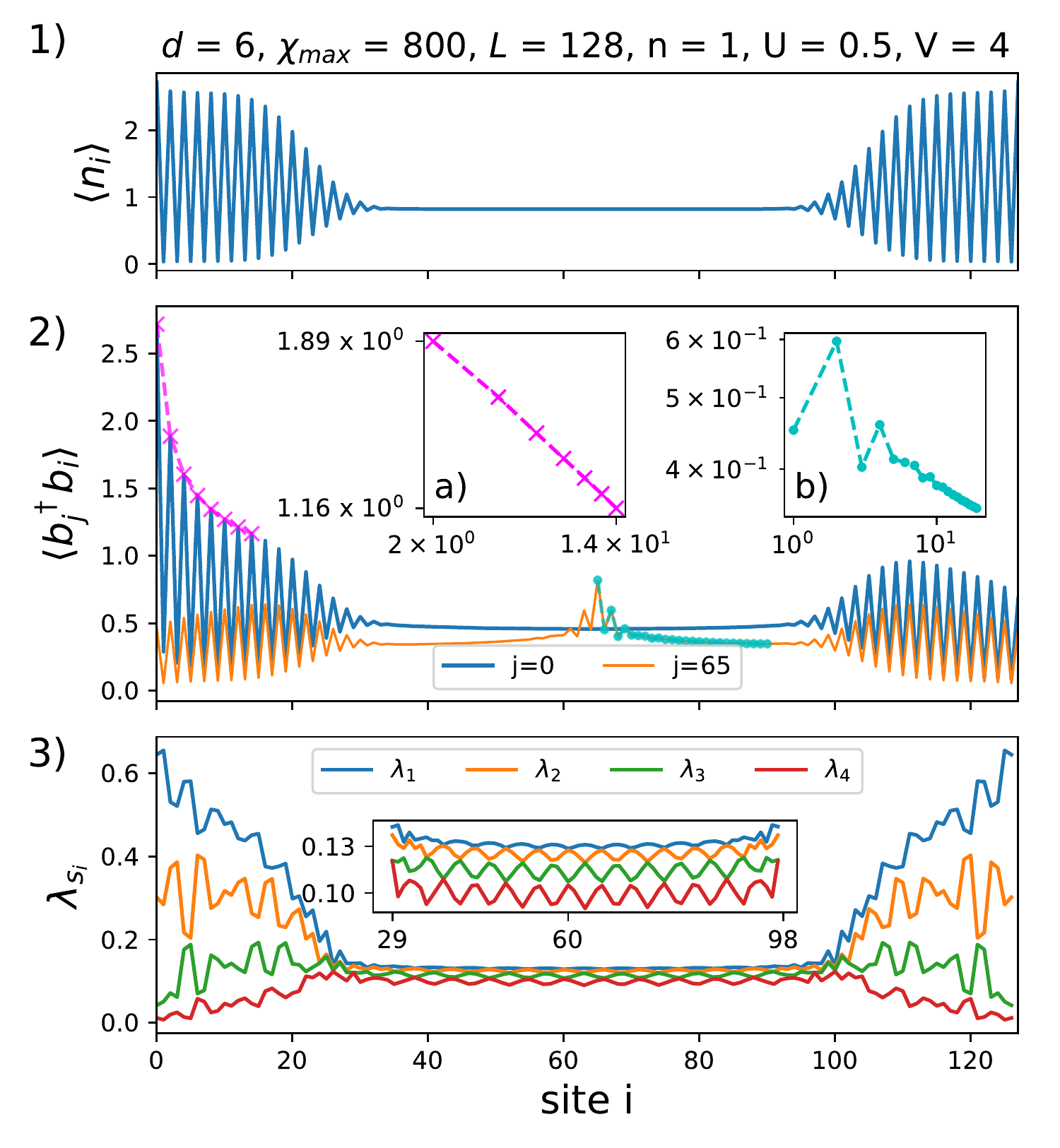}
\caption{
Main characteristics of the phase-separated ground state. 
Panel 1) Density profile. The system is separated into two phases described by a flat density typical to a fluid (SF phase) and a periodic structure typical to a solid (SS phase). 
The solid patterns of alternating occupation are pinned at the edges due to the use of open boundary conditions, leaving the superfluid uniform density in the middle. 
The volume occupied by each of the phases depends on the filling $n$, which here is $n=1$. 
Panel 2) Off-diagonal single-particle correlation function $\braket{b_j^\dagger b_i}$ in SF and SS phases.
Slow power-law decay [seen as straight lines in a log-log plot in insets 2(a) and 2(b)] allows the system to be coherent at distances larger than the lattice spacing which is a one-dimensional analog of Bose-Einstein condensation, and implies that both phases are superfluid.
3) The entanglement spectrum shows different periodicities in the two different phases.
The first four largest elements of the entanglement spectrum $\{\lambda_i\}$ are plotted in descending order $\lambda_1 \geq \lambda_2 \geq \lambda_3 \geq \lambda_4$.}
\label{fig:SS_SF_separation}
\end{figure}

For large nearest-neighbor interactions and weak on-site interactions ($(U,V) \sim (0.5,4)$), we observe a phase separation (PS) into a supersolid (SS) and superfluid (SF) region that manifests in the characteristic onset of a density wave order starting from the boundaries of the system (see \cref{fig:SS_SF_separation} 1)).
The superfluid region shows a power-law decay of $\CSF$ and a uniform density, whereas the supersolid region features staggered local densities with a simultaneous presence of coherence as indicated by the power-law decay of $\CSF$ (see \cref{fig:SS_SF_separation} 2)).

\begin{figure}
\centering
\includegraphics[width=.49\textwidth]{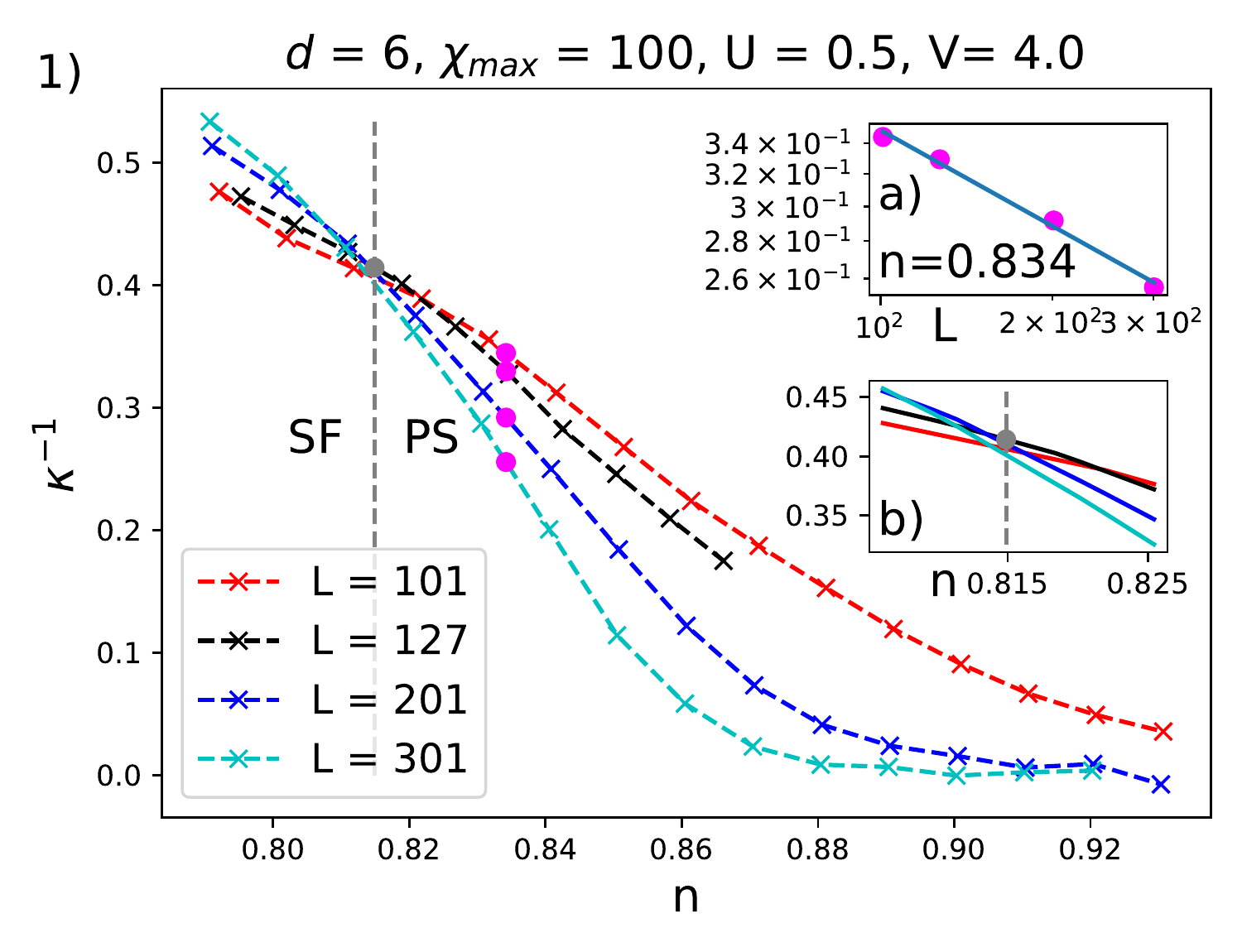}
\includegraphics[width=.49\textwidth]{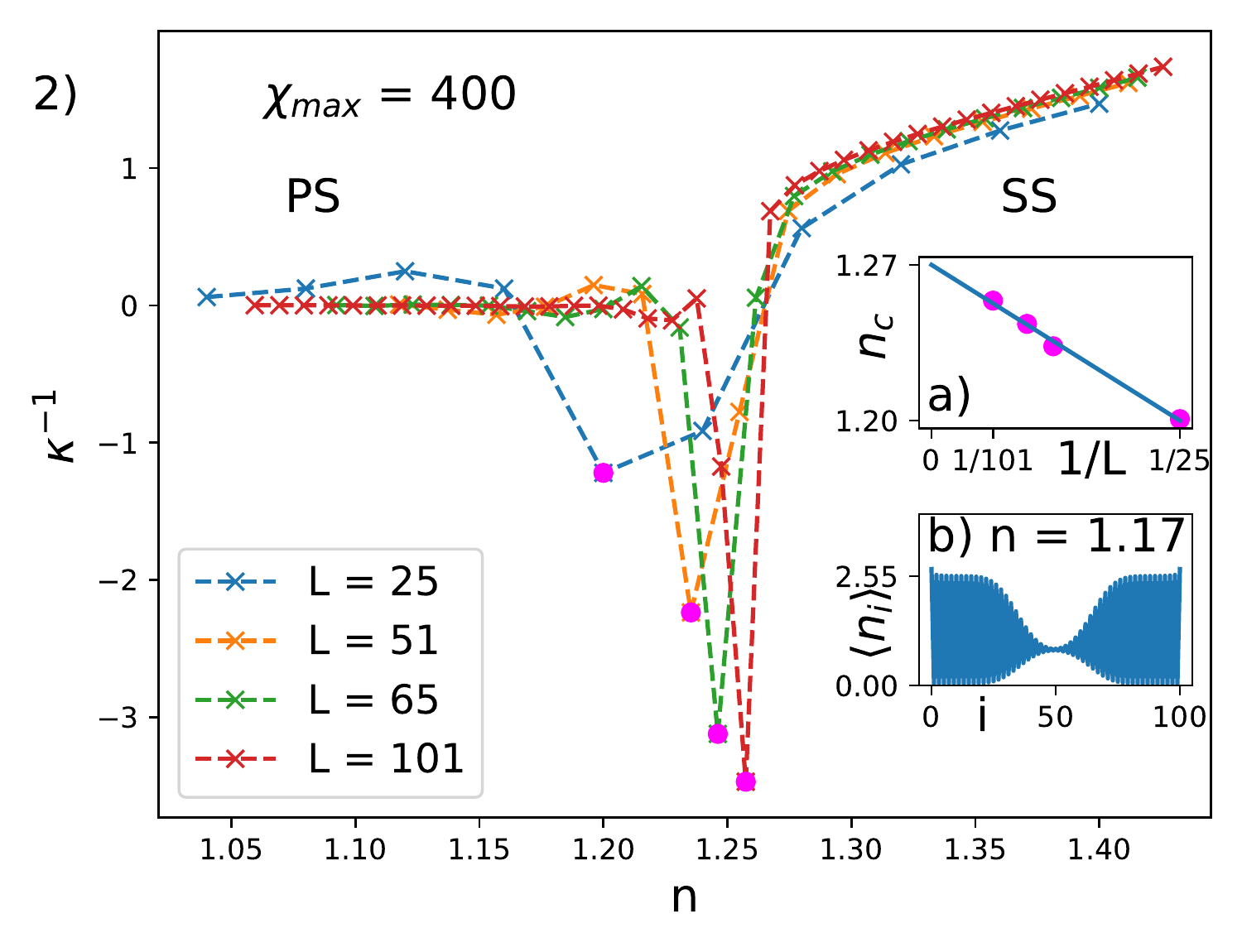}
\caption{
Finite-size study of the inverse compressibility $\kappa^{-1}$, Eq.~(\ref{eq:kappam1}), as a function of the filling $n$ for $(U,V) = (0.5,4)$ as in fig.~\ref{fig:SS_SF_separation}.
Vanishing thermodynamic value of $\kappa^{-1}$ \cref{eq:kappam1} signals instability towards phase separation. 1) SF-PS transition: The transition point is estimated at critical filling $n^\text{SF-PS}_c = 0.815$ defined as the position of the intersection of lines corresponding to different system sizes.
In the phase-separated region, $n>n_c$ region, the value of the inverse compressibility $\kappa^{-1}$ is lowered as the system size is increased (lines correspond to $L=101; 127; 201; 301$, from top to bottom) and vanishes in the thermodynamic limit.
Inset~(1a): example power-law decay of the inverse compressibility $\kappa^{-1}$ as a function of system size $L$ in the phase-separated regime, $n>n_c$ 
Inset~(1b): Zoom-in on the intersection. 2) PS-SS transition: Dependence of the inverse compressibility on filling $n$ is scaled with the system size $L$ leading to the estimated value for the critical density equal to $n^\text{PS-SS}_c = 1.27$ (Inset~(2a)). Inset~(2b): Example density in PS state close to the transition to SS. Note that in the solid part the average density $n\approx 2.55/2$ matches the critical filling $n^\text{PS-SS}_c$.}
\label{d2E-S_f}
\end{figure}

A phase transition happens if the equation of state, describing the dependence of the chemical potential on the density, has two minima. The single-minimum scenario converts to the two-minima one when an inflection point, $d\mu/dn = 0$ appears. Indeed, it is known that the divergence in compressibility leads to a phase separation \cite{Grilli1991,Emery1993,Misawa2014} and this criteria has been used to locate its position numerically. Thus, occurrence of the phase separation can be understood as a mechanical instability of the system, signaled by a vanishing inverse compressibility \cite{Emery1990,Ammon1995,Coulthard,Moreno2010}
\begin{equation}
\label{eq:kappam1}
\kappa^{-1} = n^2 \frac{\partial^2 \mathcal{E}}{\partial n^2} \approx n^2 \frac{\mathcal{E}(n +\Delta n) + \mathcal{E}(n - \Delta n) - 2 \mathcal{E}(n)}{\Delta n^2}
\end{equation}
where $\mathcal{E} = E_0/L$ is the ground state energy density and $n = \sum_i \braket{n_i}/L$ the average particle density. For these calculations, we fix $L$ and vary $n=N/L$ in an equidistant manner $N\in\mathbb N$, such that $\Delta n = (N_1 - N_0)/L$ for different fillings. The system becomes mechanically unstable and phase separation occures when the compressibility becomes infinite (or $\kappa^{-1}=0$) \cite{Moreno2010}. We show that this is exactly the case and report the finite-size scaling of the SF-PS transition in fig.~\ref{d2E-S_f}. We estimate the transition point at the crossover of for different finite system sizes as $n_{c}^\text{SF-PS} \approx 0.815$ (see~\cref{d2E-S_f} inset 1b) for a detailed view). For larger fillings, $n>n_c$, the inverse compressibility $\kappa^{-1}$ tends towards zero in the thermodynamic limit (see~\cref{d2E-S_f} inset 1a)), signaling spinodal decomposition leading to the phase-separated ground states for intermittent fillings.
We estimate the critical filling as $n_{c}^\text{PS-SS} \approx 1.27$ from extrapolating the points for which the second derivative changes abruptly (see~\cref{d2E-S_f} a)).
This filling coincides with the average density of the SS part in the PS configuration $n\approx 2.55/2$ in the vicinity of PS transition, as shown in~\cref{d2E-S_f} inset 2b). 
To rule out artifacts from the restricted local Hilbert space dimension, we achieve consistent results for maximal local occupation number $d=4,6,9$ and found no significant differences between $d=6$ and $d=9$ (see \cite{suppl}). As a compromise between performance and accuracy, we fixed $d=6$ for all presented calculations.

The surface energy between SS and SF phases is minimized in configurations with only two domains. Open boundary conditions, employed in DMRG calculations, pin the solid region to the edges while the superfluid one is observed in the center [see fig.~\ref{fig:SS_SF_separation}].
The solid region appears at random positions within the unit cell in iDMRG calculations, where unit cells are repeated periodically, as one would expect in a phase-separated ground state.

\begin{figure}
\centering
\includegraphics[width=.49\textwidth]{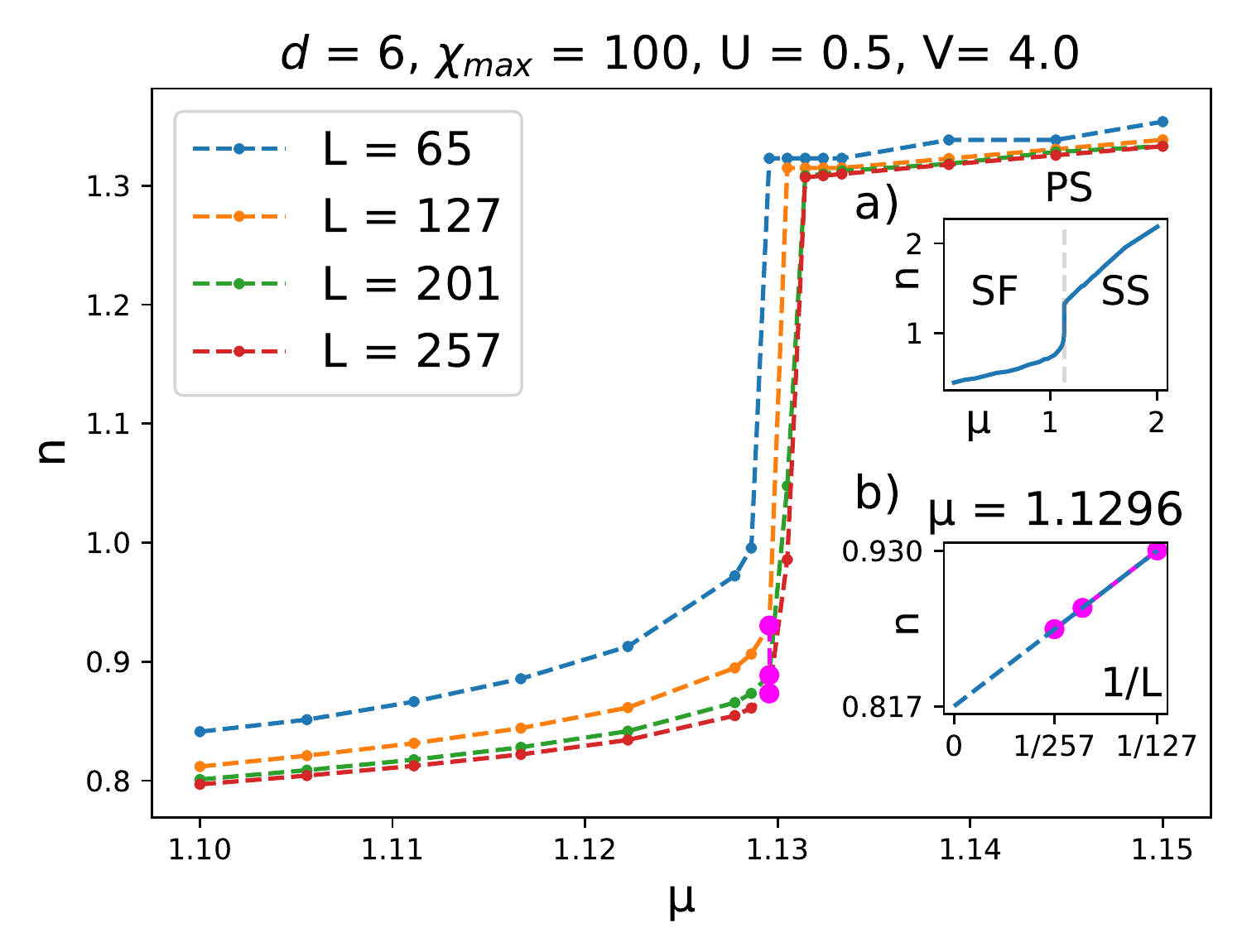}
\caption{Filling $n$ versus chemical potential $\mu$ for finite chains of length $L$ with OBC and no explicit $U(1)$ symmetry. The apparent discontinuity at roughly $\mu \approx 1.13$ signals spinodal decomposition for fillings between $n=0.817$ and $1.31$. 
Insets (a) Wider range in $\mu$ showing the extent of the SS phase, 
(b) Finite-size extrapolation for the filling values in the main plot at $\mu = 1.1296$ to estimate the lower critical filling $n_c=0.817$.
}
\label{fig:f_mu_n-max-5}
\end{figure}

\begin{figure}
\centering
\includegraphics[width=.47\textwidth]{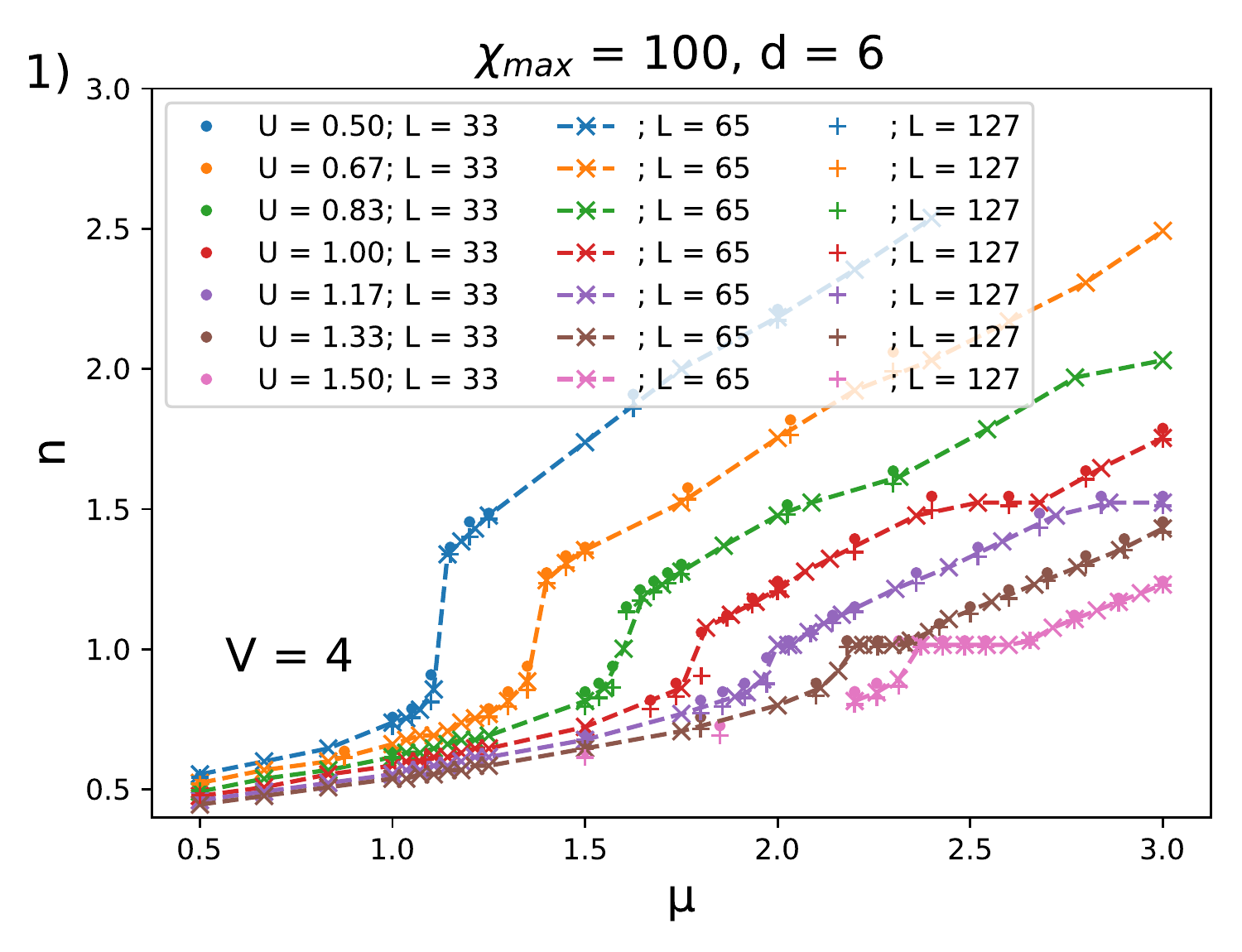}
\includegraphics[width=.47\textwidth]{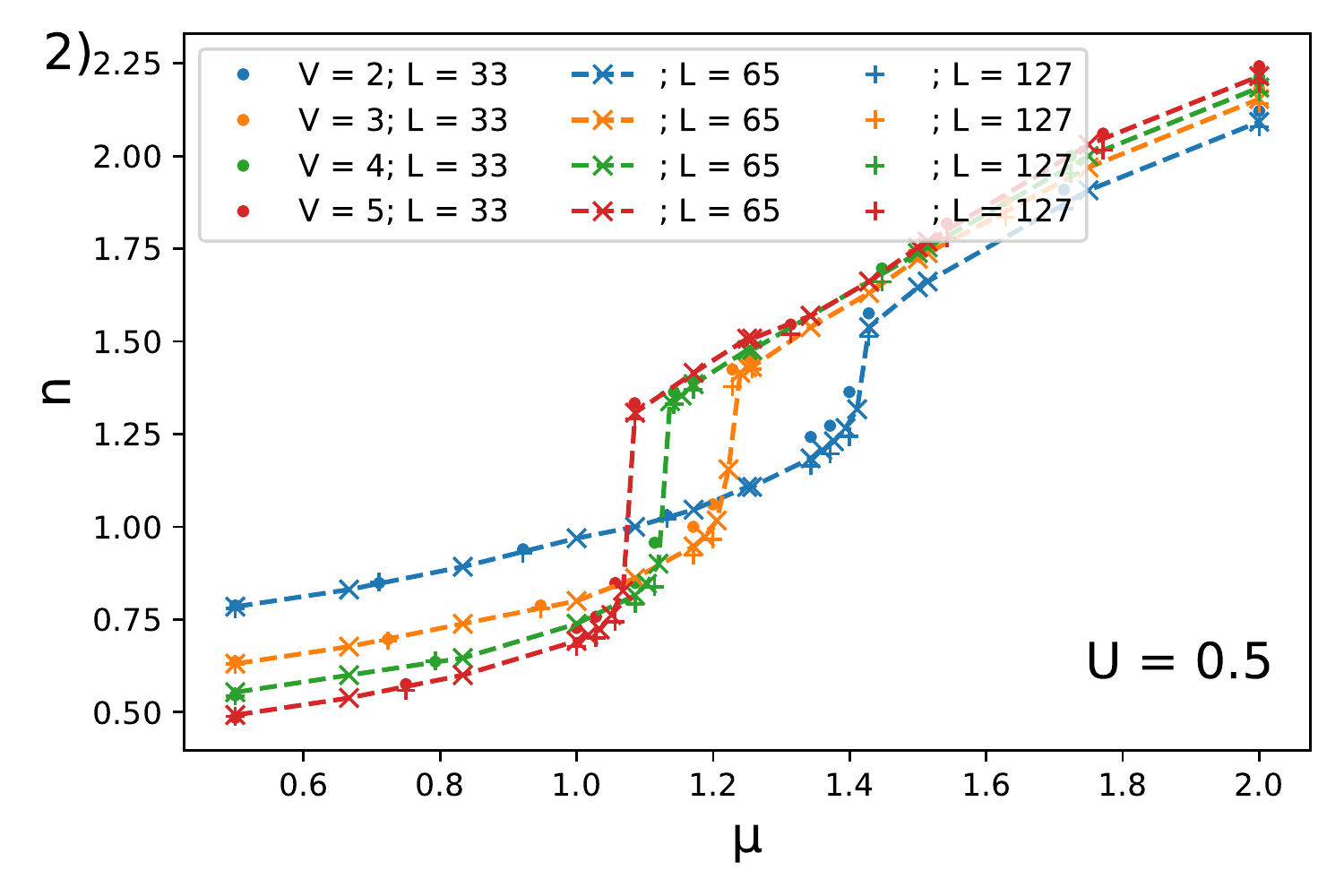}
\caption{Dependence of filling $n$ on the chemical potential $\mu$ for different system sizes $L$ and different values of $U$
The phase separation is seen as a discontinuity in $n(\mu)$. 1) Fixing $V=4$ we alter $U$ and see that beyond $U_c \approx 1$ the systems form a CDW, seen by the plateaus at filling $n=1$ (lines are ordered in increasing order of $U$ from the top line to the bottom one).
2) For smaller $V$, PS occurs at higher fillings such that it is not present in the $n=1$ phase diagram for small $V$ anymore (lines are ordered in increasing order of $U$ from the top lines to the bottom ones).
}
\label{fig:n-mu-U}
\end{figure}

An alternative way to narrow down the appearance of phase separation is via altering the chemical potential $\mu := \partial \mathcal{E} / \partial n$ (note that $\kappa = n^{-2} \partial n / \partial \mu$). In fig.~\ref{fig:f_mu_n-max-5} we show the filling $n(\mu)$ obtained with open boundary conditions for finite chains as we vary the chemical potential $\mu$. Notably, we observe a discontinuity at $\mu_c \approx 1.13$, exactly at the point where the compressibility $\kappa$ becomes infinite. We extrapolate the critical fillings to be between $n_{c} \in [0.82,1.31]$. This is in agreement with the densities we obtained in the previous calculation.
We show in fig.~\ref{fig:n-mu-U} how the dependence $n(\mu)$ changes if we alter $(U,V)$. In fig.~\ref{fig:n-mu-U}(1) discontinuities in $n(\mu)$ are clearly visible, signaling formation of a PS state below a critical $U_c(V=4) \approx 1$. For larger values of $U$, the system forms a CDW phase at commensurate fillings, signaled by the formation of plateaus with constant $n(\mu)$ (i.e. $n=1$ here).
From \cref{fig:n-mu-U} (2) we observe that the phase separation occurs for larger average densities $n>1$ if the nearest neighbor interaction is weak. 
Therefore, the reported effects go beyond the usual commensurate effects between lattice geometry and average density.

\section{Spatial oscillations in SF phase}
\label{sec:oscillations}
\begin{figure}
\centering
\includegraphics[width=.49\textwidth]{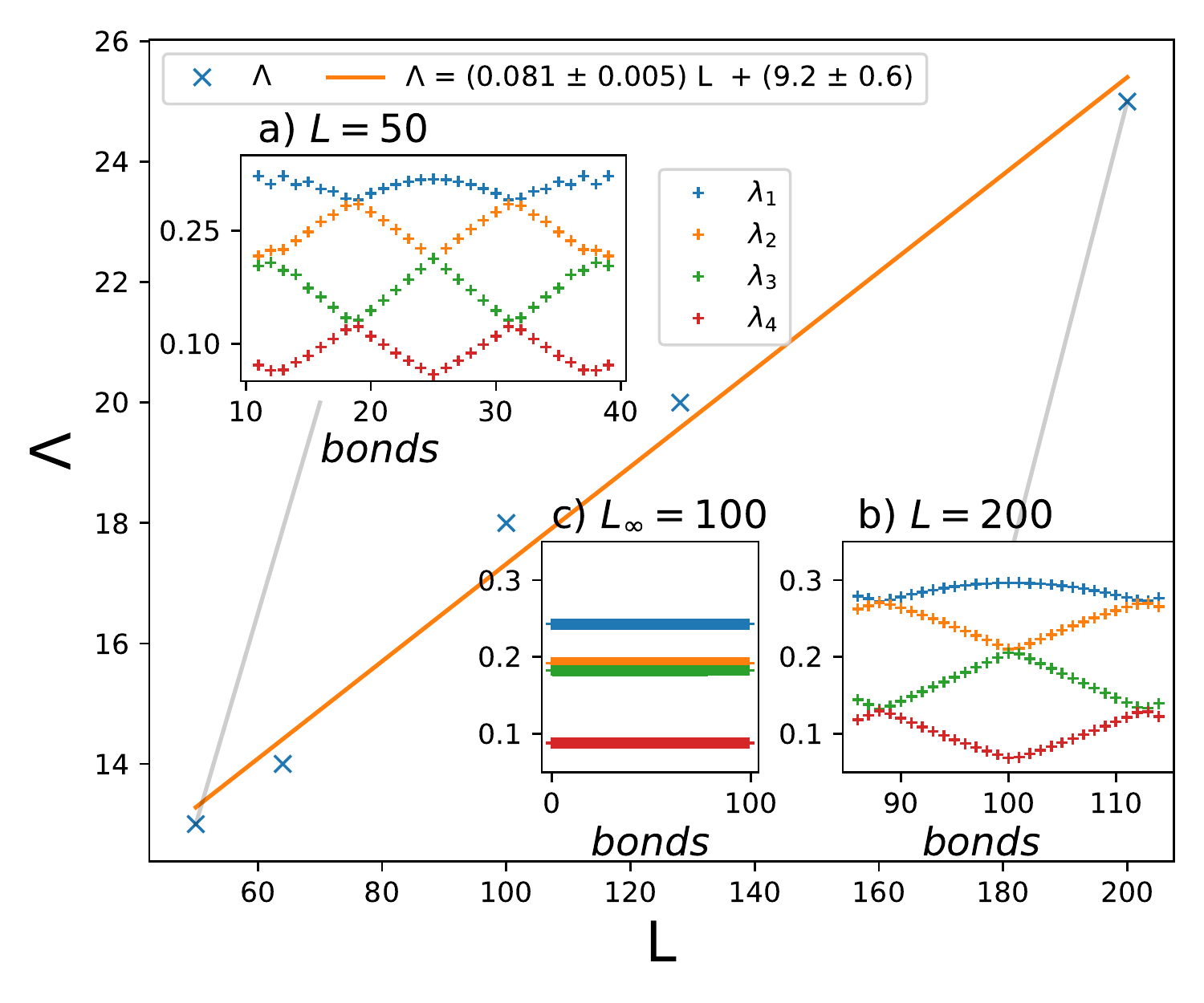}
\caption{
Spatial period $\Lambda$ taken from the Entanglement spectrum (ES) at $(U,V) = (0.5,3)$ with OBC and $\chimax = 100$, $d = 6$ as a function of the system size $L$. This state was reported as SS in Ref.~\cite{Deng2011}, but it is actually an SF. Further, the spatial oscillations vanish in the thermodynamic limit as the spatial period grows linearly in system size. However, we show that for incommensurate fillings like in \cref{fig:OBC_f-0-77,fig:iDMRG_f-0-65_Csl-Ssl-ES} this kind of oscillations do survive in the thermodynamic limit.
a,b) ES $\lambda_{s_i}$ for bonds at the center of a system of length $L=50$ and $L=200$, respectively. c) ES for iDMRG simulation in the thermodynamic limit with a unit cell size $L_\infty = 100$ showing no spatial oscillations.
}
\label{fig:reproduce_santos}
\end{figure}

\begin{figure}
\centering
\includegraphics[width=.49\textwidth]{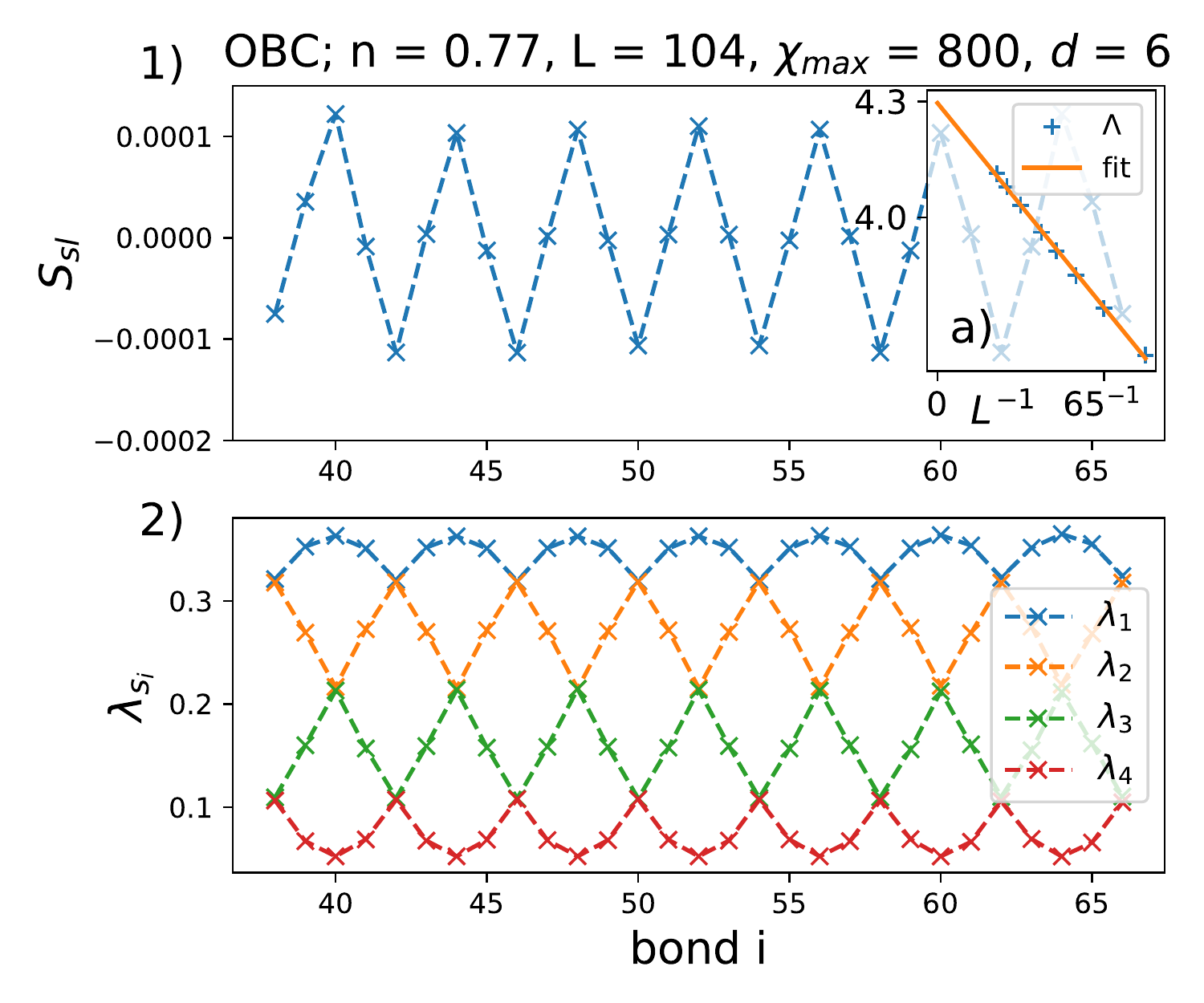}
\caption{Entanglement properties of the superfluid bulk for filling $n=0.77$ and $(U,V) = (0.5,4)$ close to phase-separation. 1) Oscillatory sub-leading part $S_\text{sl}$ of the R´enyi-2 entropy $S_2$ in \cref{Renyi_scaling}. 2) The four largest squared Schmidt coefficients $\lambda_{s_i}$ shown spatially along the bonds. The inset 1a) shows spatial frequencies from Fourier analysis of $\lambda_{s_i}$. The extrapolation yields a spatial period $\Lambda = 4.3$ in the thermodynamic limit.}
\label{fig:OBC_f-0-77}
\end{figure}

\begin{figure}
\centering
\includegraphics[width=.49\textwidth]{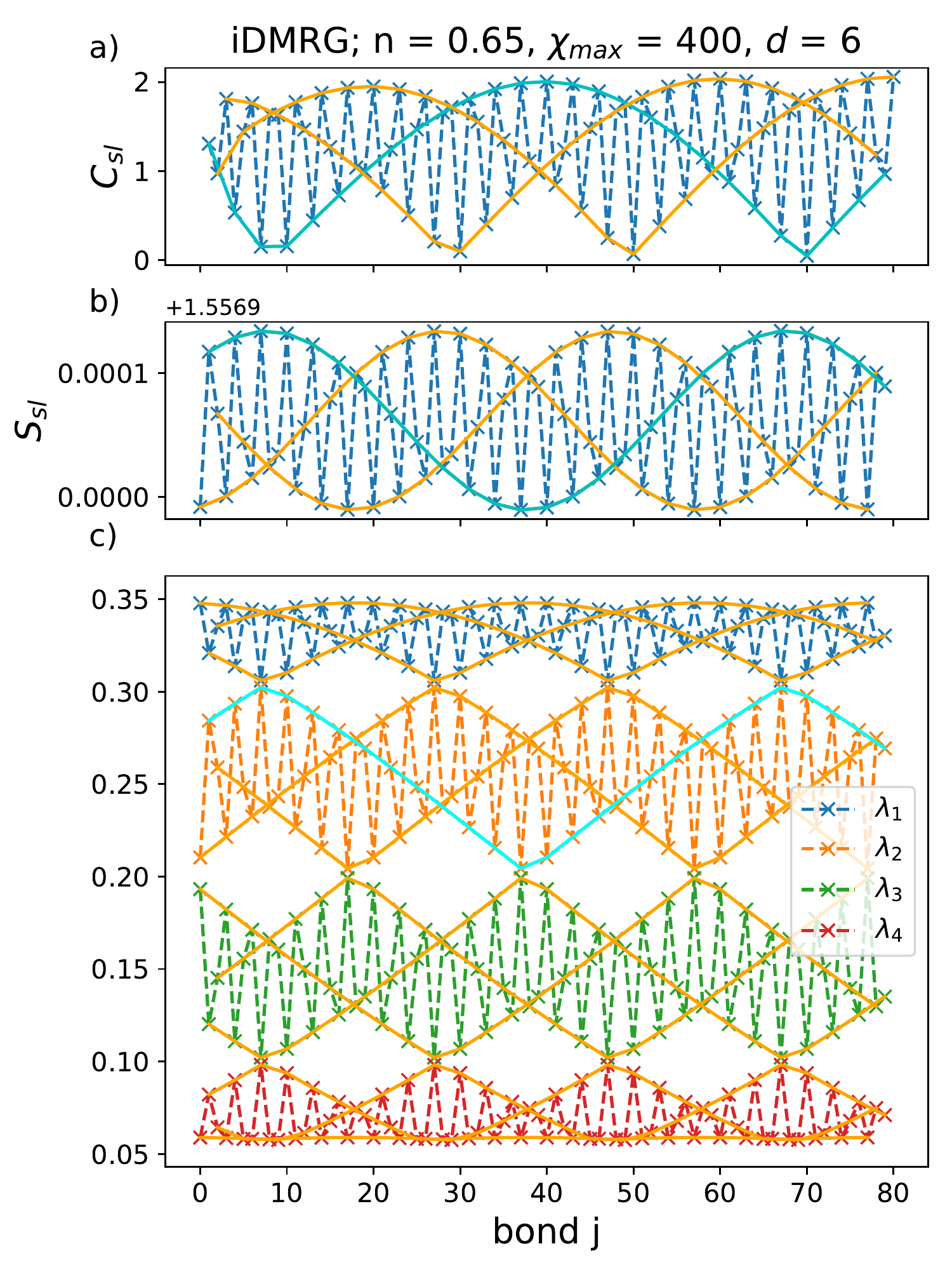}
\caption{
a) Sub-leading part $C_\text{sl}$ containing the oscillatory part of the string-order correlator $\Cst$ \cref{eq:Cs}. b) Oscillatory part of the R´enyi-2 entropy $S_2$. c) The spatial dependence of the four largest squared Schmidt coefficients $\lambda_{s_i}^2$ on the bonds of the MPS. For all a),b) and c): The orange lines are guides to the eye with every third data point plotted to highlight the envelope. We highlight one line in cyan color in all three subplots as a guide to the eye, making it apparent that the period is the same in all three quantities.}
\label{fig:iDMRG_f-0-65_Csl-Ssl-ES}
\end{figure}

\begin{figure}
\centering
\includegraphics[width=.48\textwidth]{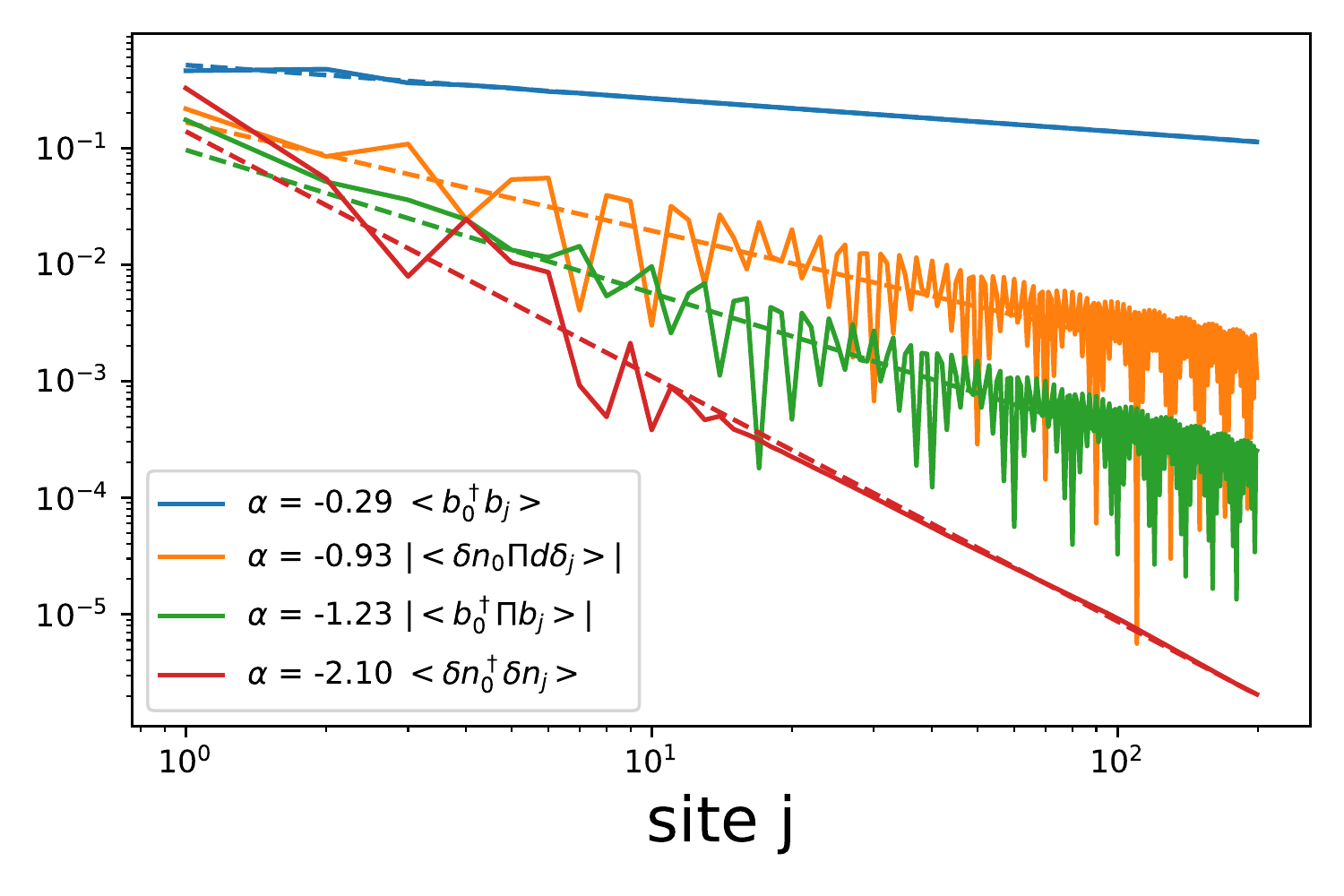}
\caption{Comparison of different common correlators for the same state as in \cref{fig:iDMRG_f-0-65_Csl-Ssl-ES} in the SF phase with oscillating ES. 
We denote the non-local string term as $\Pi = \exp\left(-i \pi \sum_{0 \leq l < j} \delta n_l\right)$ with $\delta n_l =  n_l - n$. 
We see that only the correlators with this string term show the oscillations matching the ES. The critical exponent $\alpha$ is obtained from linear fitting the respective correlator in a double logarithmic scale (dotted lines in corresponding colors).}
\label{fig:iDMRG_correlators_f-065}
\end{figure}

For the homogeneous superfluid at fillings below the phase-separated phase, enhanced spatial oscillations appear in the entanglement spectrum (ES) and other observables (see \cref{fig:reproduce_santos,fig:OBC_f-0-77,fig:iDMRG_f-0-65_Csl-Ssl-ES,fig:iDMRG_correlators_f-065}).
These signatures are present in SF states for weak on-site interactions $U$ over a broad range of $V$.
Such oscillatory patterns were reported earlier in Ref.~\cite{Deng2011} at $(U,V) = (0.5,3)$ and $n=1$.
However, for the superfluid at integer fillings, we observe the absence of oscillations in the thermodynamic limit such that they cannot be linked to a bulk feature of the given phase and must be related to finite-size effects, instead.
We demonstrate this in \cref{fig:reproduce_santos}, where we show the spatial period of the oscillatory patterns in the entanglement spectrum (examples thereof are visible in insets a) and b)) as a function of the system size, for which we observe a linear increase, i.e. a vanishing frequency for $L\rightarrow\infty$. 
This is in agreement with iDMRG simulations (thereby directly approximating the ground state in the thermodynamic limit), for which we do not find oscillations at all. It shall also be noted that the system is in a superfluid phase for these parameters, and not a supersolid phase as claimed in \cite{Deng2011}.
The commensurate scenario at $n=1$ is in strong contrast to incommensurate fillings, for which oscillations in the entanglement spectrum are a robust feature of the bulk.

In the inset of \cref{fig:OBC_f-0-77}~a), we display a finite-size extrapolation of the spatial period, which is extracted from the leading frequency in the Fourier transform of the oscillatory part of the entanglement spectrum (see \cref{fig:OBC_f-0-77}~2))).
Notably, $\Lambda(1/L\rightarrow0)\approx4.3$ assumes a finite value in the thermodynamic limit.

The oscillations of the entanglement spectrum cannot be detected by standard local observables and two-body correlations, but, interestingly, they appear prominently in non-local observables like the string-order correlator, for which the long-range power-law decay is modulated by oscillations of the same frequency (c.f. \cref{fig:iDMRG_f-0-65_Csl-Ssl-ES}~a)).
In order to extract the oscillatory part of $C_\text{sl}$ we fit it with a power-law decay $\Cst(i,j) = c/|i-j|^\alpha C_\text{sl}$ and divide the correlator by the envelope $c/|i-j|^\alpha$ to show the remaining oscillatory part.
In contrast, common correlators without the non-local string term do not show this oscillatory behavior as depicted in  \cref{fig:iDMRG_correlators_f-065}. All of the correlators here decay algebraically, in particular the ones involving the non-local string term, indicating lack of long range trivial and topological order, but power-law correlations. The correlators with string term exhibit oscillations in the tails reminding us of the oscillations of the ES.

A complementary way to resolve these spatial oscillations is given by the R´enyi entropy in \cref{eq:renyi_entropy}, accessible in experiments for the special case $\alpha=2$.
$S_2(\ell)$ depends on the purity $\rho^2(\ell)$ for a lattice block of size $\ell$, which can be detected in the framework of trapped ions through quantum state tomography~\cite{linke2018} or through direct measurement of the quantum purity~\cite{islam2015}.

The asymptotic decay of the R´enyi entropies $S_\alpha(\ell)$ for critical systems is well-known~\cite{calabrese2004,calabrese2009} and given by Eq.~(\ref{Renyi_scaling}).
The leading contribution to $S_\alpha(\ell)$ is proportional to $\ln\left( d[\ell|L] \right)$ and describes a universal scaling law with prefactor factor $c$ called the central charge of the conformal field theory (in a fermionic system, it is equal to the number of Fermi points).
An additional factor $b$ distinguishes the case of periodic ($b=1$) and open ($b=2$) boundary conditions and $\gamma$ constitutes a non-universal constant.
Subleading terms are denoted by $S_{\rm sl}$ and, in general, oscillate in space.
The subtle oscillations of the entanglement spectrum are obviously carried over to the subleading terms of the R´enyi entropies, which we present in \cref{fig:iDMRG_f-0-65_Csl-Ssl-ES,fig:OBC_f-0-77}.

We note that we observe the same properties for the homogeneous SS \textit{above} the filling for phase separation. The main difference is that for this SS, there is a spatial solid pattern in the density. Taking this into account, the remaining properties are the same as is shown in the SM \cite{suppl}.

Overall, the spatial oscillations in the entanglement spectrum, that can be uncovered by looking at the string order correlators or R´enyi entropies, are a clear manifestation of a broken translational symmetry. This point is further strengthened by simulations with iDMRG: Upon choosing a suitable unit cell size, iDMRG can converge, as is the case in \cref{fig:iDMRG_f-0-65_Csl-Ssl-ES}, and yields results in agreement with the bulk of finite size simulations. For unit cell sizes incommensurate with the spatial period, iDMRG has trouble converging. We show further details about this in the SM \cite{suppl}. This feature distinguishes the superfluid phase under investigation from the superfluid phase at filling $n=1$. In the following, we convince ourselves that this phase is still well described by Luttinger liquid theory and shows the same critical finite-size scaling behavior.

\section{Luttinger liquid description}
\label{sec:luttinger}
Long-range properties of gapless one-dimensional systems are well captured by the Luttinger liquid theory and are governed by the Luttinger parameter $K$. 
This theory is based on using an effective low-energy Hamiltonian and can be used to calculate the small-momentum and long-range behavior of the correlation functions. 
The Luttinger theory, being an effective one, takes the Luttinger parameter as an input and an independent calculation is needed to relate the value of $K$ to the microscopic parameters of the lattice model.
In the following, we use two independent ways to calculate $K$, which is useful for the characterization of the system properties. Furthermore, it serves as a stringent test for the internal consistency of the numerics.

\begin{figure}
\centering
\includegraphics[width=.49\textwidth]{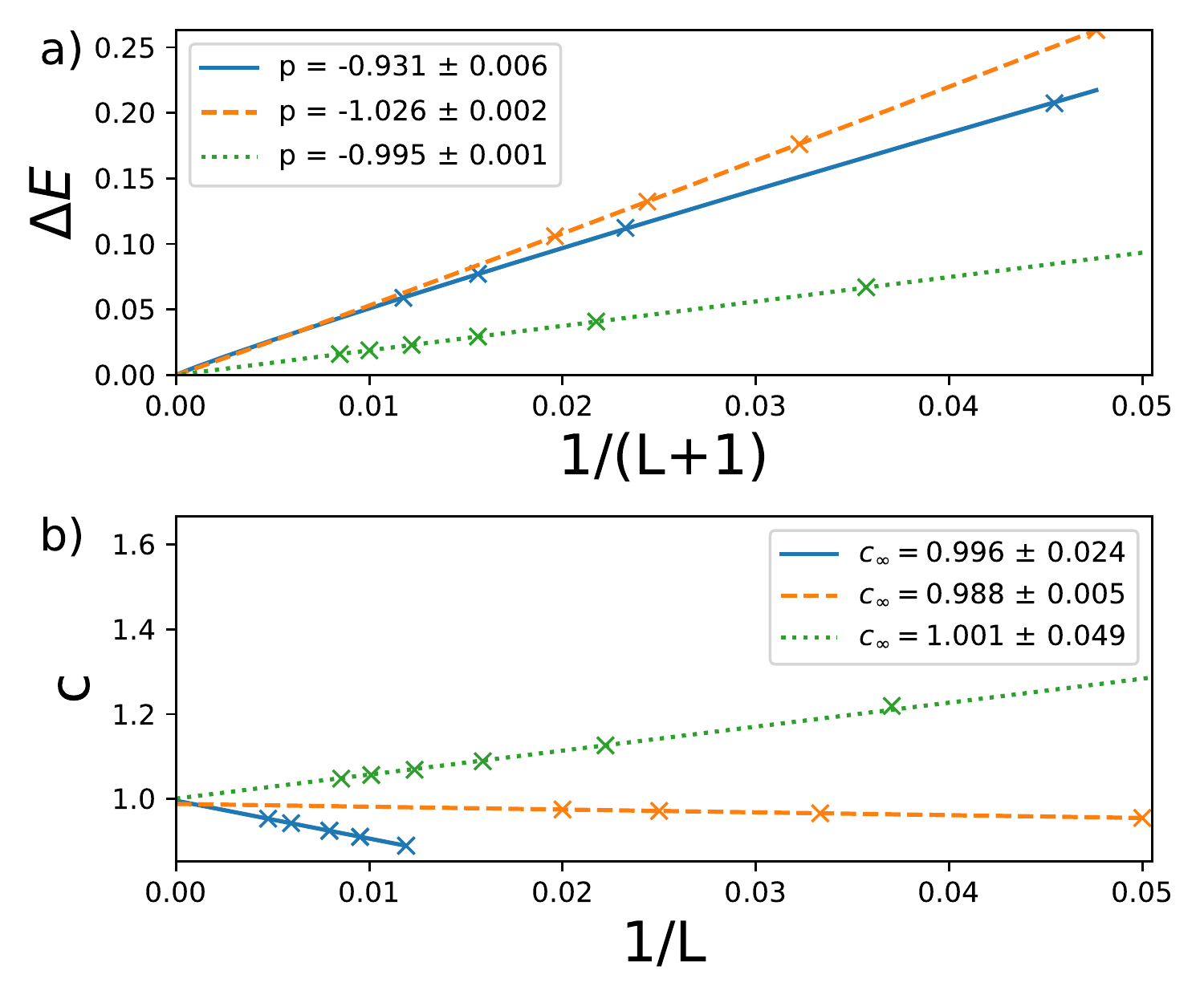}
\caption{Critical scaling for a superfluid and supersolid states with and without spatial oscillations.
Solid blue lines: SF at $(U,V) = (0.5,4)$ with $n=0.62$ and spatial oscillations. Dashed orange lines: SF at $(U,V) = (0.5,0.5)$ at integer filling without spatial oscillations. Dotted green lines: SS at $(U,V) = (0.5,4)$ with $n=1.333$ and spatial oscillations. Despite very different spatial features, all states seem to be well described within the same field theoretic description. a) Finite-size scaling of the level splitting $\Delta E = E_1 - E_0$ vanishing in the thermodynamic limit and yielding the critical exponent $p$ from $\Delta E \propto 1/(L+1)^p$, roughly matching the expected $p_\text{LL}=-1$ for a Luttinger liquid. b) Central charge $c$ \cref{fig:central_charge_entropy} extrapolated to the thermodynamic limit $c_\infty$ from $c = c_\infty + \text{const}./L$ matching the central charge $c_\text{LL} = 1$ for a spinless Luttinger liquid.}
\label{fig:deltaE_c}
\end{figure}

We check various other quantities and compare them with known parameters for an SF without spatial oscillations.
Within the Luttinger liquid description, the lowest-lying excitation spectrum is considered to be linear in momentum $k$, i.e. $E(k) = \hbar k v_s$, where $v_s$ is the speed of sound.
Furthermore, the speed of sound is related to the compressibility through $mv_s^2 = n \partial \mu/\partial n = (n\kappa)^{-1}$\cite{LLvolumeIX}.
In a finite-size (open boundary) system of size $L$, the minimum allowed value of the momentum is inversely proportional to the length of the wire, i.e. $k_{min} = \pi/(L+1)$. 
As a result, the excitation spectrum has a level splitting
\begin{equation}
\Delta E = \frac{\pi\hbar v_s}{L+1}
\label{eq:Delta E}
\end{equation}
which vanishes in the thermodynamic limit, $L\to\infty$.
We confirm this antiproportional scaling $\Delta E \propto (L+1)^p$ in fig.~\ref{fig:deltaE_c}(a) by a fit of the critical exponent $p = -0.931 \pm 0.006$, which is consistent with the expected value of $p_\text{LL} = -1$.

We further extract the central charge $c$ from the von Neumann entropy (R´enyi entropy in the limit $\alpha \rightarrow 1$)
\begin{equation}
\label{fig:central_charge_entropy}
  S_1 = \frac{c}{6} \log(d[\ell|L])
\end{equation}
for which we obtain an extrapolated value of $c(L\rightarrow\infty) \approx 0.99$ throughout the superfluid phases, which is in perfect agreement with the predicted result $c_\text{LL} = 1$ for a spinless Luttinger liquid (see~\cref{fig:deltaE_c} panel~b)).

\begin{figure}
\centering
\includegraphics[width=.49\textwidth]{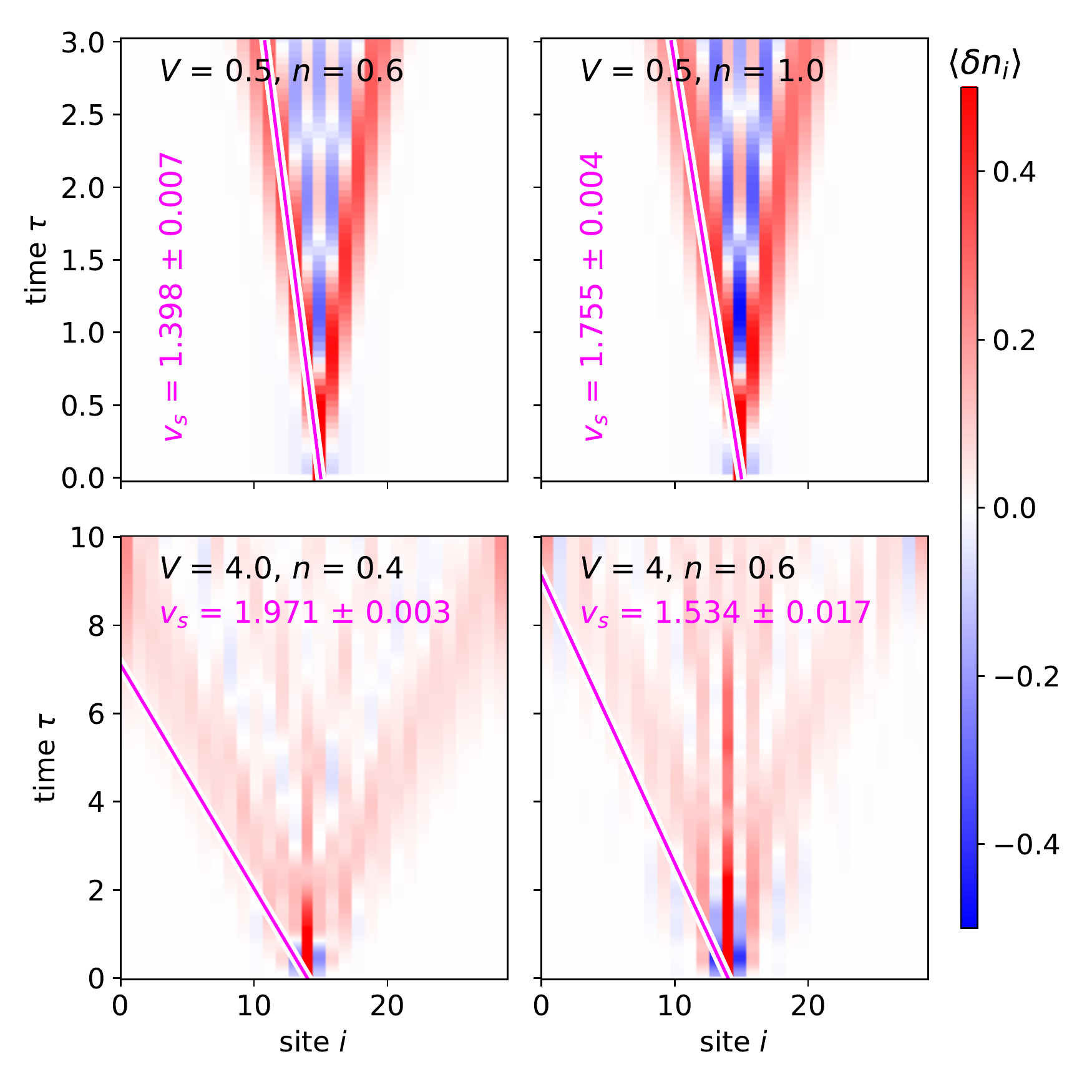}
\caption{Dynamical analysis of a local perturbation. 
We create a particle at a central site of the ground-state wave function and track the time evolution of the space-resolved density.
The propagating defect $\braket{\delta n_i} = \braket{\Psi(\tau)|n_i|\Psi(\tau)} - \braket{\Psi_0|n_i|\Psi_0}$ 
shows a typical ``light-cone'' structure, for which the red and blue coloring corresponds to positive and negative excess with respect to the average ground-state density. 
The boundary of the cone propagates with the speed of sound (highlighted in magenta), obtained by fits of the level splitting according to~\cref{eq:Delta E}. 
In all panels, we fixed $U=0.5$ and parameters $L=30$, $\chimax = 400$, and $d=6$.
}
\label{fig:lightcones}
\end{figure}

To check the validity of \cref{eq:Delta E}, we compute the speed of sound $v_s$ at four distinct points in parameter space $(V,n) \in \{(4,0.6), (4,0.4), (0.5,1), (0.5,0.6)\}$ with fixed $U=0.5$ and compare them with dynamical simulations. For this, we disturb the ground state at the middle of the chain $\ket{\delta \Psi_0} = b^\dagger_{L/2} \ket{\Psi_0}$ and compute its time evolution $\ket{\Psi(t)} = \exp\left(-i\tau H \right) \ket{\delta \Psi_0}$ via trotterization (TEBD) \cite{White2004TEBD,Vidal2004TEBD}. We then compute the density distribution at each time step and substract the ground state density, $\braket{\delta n_i} = \braket{\Psi(\tau)|n_i|\Psi(\tau)} - \braket{\Psi_0|n_i|\Psi_0}$. The resulting lightcones are displayed in \cref{fig:lightcones} and match well the overlayed fitted speed of sound from \cref{eq:Delta E} (in magenta). Further we compare these values of the speed of sound obtained via $v_s = 1/\left( \hbar \pi n^2 \kappa K \right)$, incorporating the Luttinger parameter $K$ from~\cref{eq:KfromS} and the compressibility $\kappa$ from~\cref{eq:kappam1}. We extrapolate results to the thermodynamic limit and find a reasonable agreement within $18\%, 14\%, 1\%$ and $9\%$, again for $(V,n) \in \{(4,0.6), (4,0.4), (0.5,1), (0.5,0.6)\}$ and fixed $U=0.5$, respectively. This is a stringent test of the internal consistency of the method, as thermodynamic relation between the compressibility obtained from equation of state and the speed of sound is tested.

In the following, we rely on Luttinger liquid theory to describe the asymptotic behavior of correlation functions.
To do so, we employ an abelian bosonization analysis~\cite{Giamarchi1992},
\begin{align}
b^\dag_x\rightarrow\psi^\dag(x)\sim\sum_{m=-\infty}^{\infty}{\rm e}^{2\pi im(nx + \phi) + i\theta(x)}
\label{eq:bosonization}
\end{align}
in which $[\phi(x),\partial_{x'}\theta(x')]=i\delta(x-x')$ satisfy canonical commutation relations.
Here the $\sim$ symbol denotes equality up to a prefactor, which depends on the momentum cutoff employed to derive the low-energy description~\cite{Cazalilla2004}.
The low-energy effective Hamiltonian of the extended Bose Hubbard model in 1D results to
\begin{align}
H = \frac1{2\pi}\int{\rm d}x\left(uK(\partial_x\phi)^2+\frac uK(\partial_x\theta)^2\right) + \mathcal O_{\rm sg}
\end{align}
in which $\mathcal O_{\rm sg}$ denotes additional sine-Gordon type operators as a result of the density-density interactions which are responsible for the opening of energy gaps (e.g. in the MI and HI phase).
For the characterization of the superfluid phase, these operators are irrelevant and can be disregarded.

The local density is given by
\begin{align}
n_x\rightarrow\rho(x)=\left(n+\partial_x\phi(x)\right)\sum_l{\rm e}^{2\pi il(n x+\phi(x))}
\end{align}
in which $n$ denotes the average density.
Note that the slowly oscillating contributions correspond to $l=0$, which allows one to identify $\delta\rho(x)=\rho(x)-n\approx\partial_x\phi$ as the field encoding the local density fluctuations.
This allows to approximate the argument of the $\Pi$ operator to $\sum_{x<l<x'}\delta n_l\rightarrow \phi(x')-\phi(x)$.

Correlation functions of the rescaled fields $\phi'=\phi{\sqrt K}$ and $\theta'=\theta/(\sqrt K)$ are readily obtained by a generating functional of the corresponding quantum mechanical partition function~\cite{GogolinNersesyanTsvelik} and result in the asymptotic expressions
\begin{align}
\braket{\phi(x) \phi(x')} = -\frac1{2K}\log\left(|x-x'|\right),\\
\braket{\theta(x) \theta(x')} = -\frac K{2}\log\left(|x-x'|\right).
\end{align}
By using the identity
\begin{align}
\braket{\rm e^{\rm i\sum_k b_kf(x_k)}} = \rm e^{-\frac12\sum_{k,k'}b_kb_{k'}\braket{f(x_k)f(x_{k'})}},\quad
f\in\{\phi,\theta\}
\end{align}
we arrive at the following asymptotic forms of the correlation functions, keeping only the dominant contributions
\begin{align}
\label{eq:bdb_fit}
|\braket{\psi^\dagger(x) \psi({x'})}|
&\approx
|\braket{\rm e^{i[\theta(x)-\theta(x')]}}|
\propto|x-x'|^{-K/2},\\
\label{eq:1_fit}
|\braket{\psi^\dagger(x) \Pi \psi(x')}|
&\approx
|\braket{\rm e^{i\theta(x)}\rm e^{i[\phi(x')-\phi(x)]}\rm e^{-i\theta(x')}}|
\nonumber\\
&\propto|x-x'|^{-1/2(K+1/K)},\\
\label{eq:2_fit}
|\braket{\delta\rho(x)\Pi\delta\rho({x'})}|
&\approx
|\braket{\partial_x\phi\rm e^{i[\phi(x')-\phi(x)]}\partial_{x'}\phi}|
\nonumber\\
&\propto|x-x'|^{-1/(2K)-2},\\
\label{eq:3_fit}
|\braket{\rho(x)\rho(x')}_{\rm conn.}|
&\approx
|\braket{\partial_x\phi\partial_{x'}\phi(x')}|
\propto|x-x'|^{-2}.
\end{align}

The Luttinger liquid predictions for the long-range asymptotic of the correlation functions are verified in fig.~\ref{fig:iDMRG_correlators_f-065} and a very good agreement is found. 
Note that the oscillations observed in \cref{fig:OBC_f-0-77} and \cref{fig:iDMRG_f-0-65_Csl-Ssl-ES} are consistent with the field theoretic description if sub-leading corrections are not neglected.

\begin{figure}
\centering
\includegraphics[width=.49\textwidth]{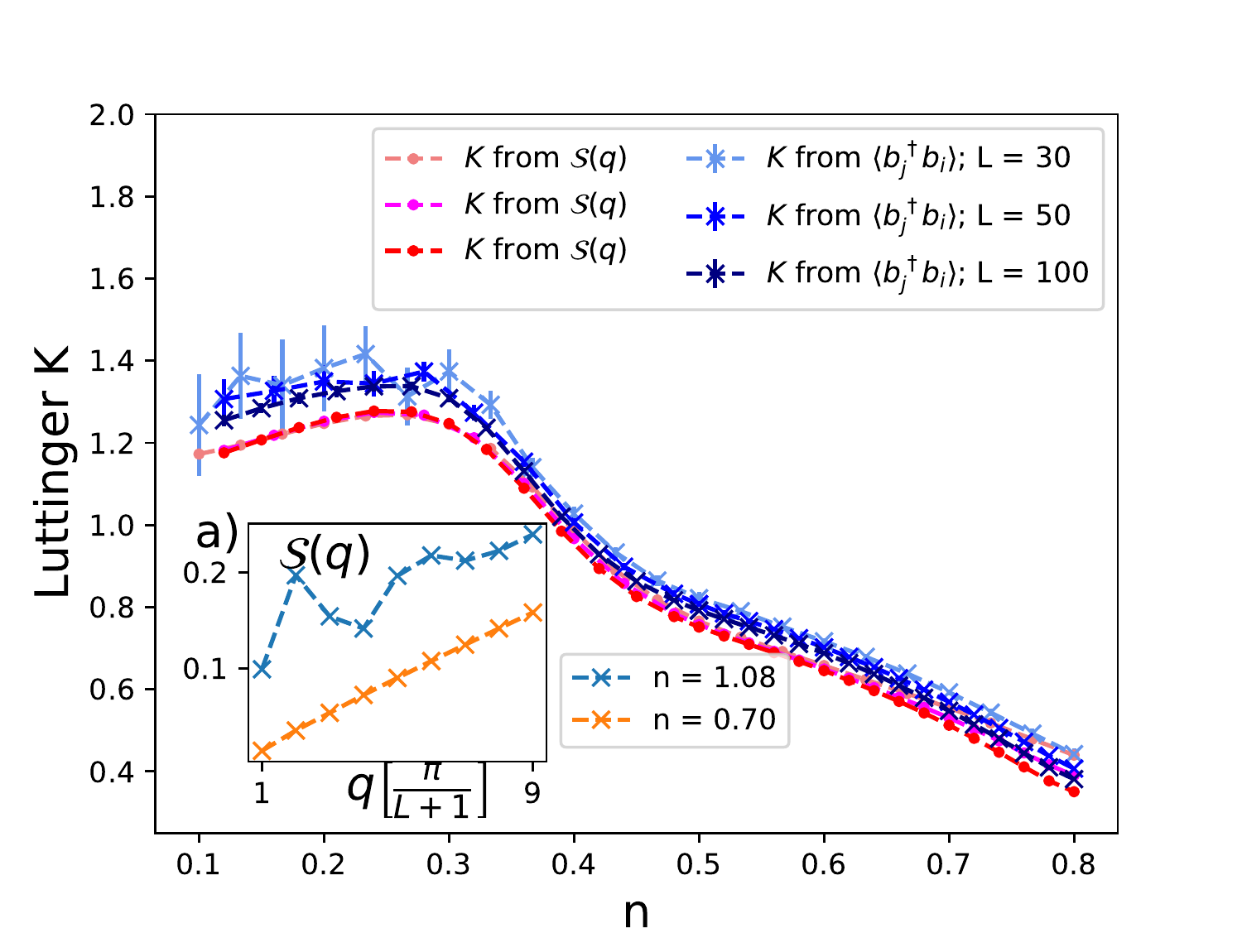}
\caption{
Luttinger parameter $K$ dependence on the filling $n$, calculated with OBC with $\chimax = 400$ and $d = 6$ at $U=0.5$ and $V=4$.
Two independent estimations are used, from the long-range asymptotic of the off-diagonal single-particle correlation function $\braket{b^\dagger_i b_j}$ \cref{eq:bdb_fit} and from the small momenta of the structure factor $\mathcal{S}(q)$ via \cref{eq:KfromS}.  
a) Structure factor $\mathcal{S}(q)$ for small momenta at different fillings. With the onset of phase separation, $\mathcal{S}(q)$ deviates from linear dependence at its origin and \cref{eq:KfromS} becomes invalid.}
\label{fig:f-K_master}
\end{figure}

Thus, the Luttinger liquid is capable of capturing correctly the long-range properties. 
At the same time, a microscopic simulation is needed to connect the parameters of the microscopic Hamiltonian to the effective parameters of the Luttinger liquid model. 
In particular, it is of great practical value to find such a relation for the Luttinger parameter $K$.
We extracted the Luttinger parameter $K$ from correlation functions in \cref{eq:bdb_fit,eq:1_fit,eq:2_fit}. 
However, we expect that oscillatory subleading terms are more important in \cref{eq:1_fit,eq:2_fit}, causing large error bars for fits of the leading order only, and we resort to a detailed comparison between the value of $K$ obtained from \cref{eq:bdb_fit,eq:KfromS} only in \cref{fig:f-K_master}.

Alternatively, the Luttinger parameter $K$ can be extracted from the slope of the linear part of the structure factor $\mathcal{S}(q) = \sum_{ij} e^{-iq(i-j)} (\braket{n_i n_j} - \braket{n_i}\braket{n_j})/(L+1)$ ~\cite{Astrakharchik2016,praga}. In the framework of the Tomonaga-Luttinger description, we can compute the Luttinger parameter via
\begin{equation}
\label{eq:KfromS}
\frac{1}{2 \pi K} = \lim_{q\rightarrow 0} \frac{\mathcal{S}(q)}{q},
\end{equation}
where $q$ and $S(q)$ depend on the system size $L$ and the boundary conditions, see \cite{Ejima2011,arik2016}. We obtain $S(q)/q$ by performing a fit of the lowest momenta where $S(q) \propto q$ is linear. 
If the lowest-lying excitation spectrum is exhausted by linear phonons, the Luttinger liquid description is applicable and the Luttinger parameter defined according to Eq.~(\ref{eq:KfromS}) is independent of the actual size of the system $L$ if it is large enough.
We see that both estimations of $K$ match well in the SF phase, as seen in \cref{fig:f-K_master}. 

The knowledge of the Luttinger parameter $K$ allows one to apply the effective description as provided by the Luttinger liquid to static and dynamic long-range properties. In particular, low-momentum behavior of the momentum distribution can be obtained as a Fourier transform of off-diagonal single-particle correlation function~(\ref{eq:1_fit}) resulting in a divergent $n(k) \propto
|k|^{1-K/2}$ behavior for $K<2$. 
That is, for all cases shown in Fig.~\ref{fig:f-K_master}, the occupation of zero-moment state diverges in the thermodynamic limit which is a reminiscence of Bose-Einstein condensation in one dimension. 
Another special value of the Luttinger parameter is $K=1/2$, below which an SF state might be sustained a unit filling as opposed to a Mott insulator which is realized for any finite height of the optical lattice\cite{zwerger2003mott,RevModPhysBloch}.
In the considered system small values of $K$ correspond to a large filling fraction $n$, further increase in $n$ leads to a phase transition.

In conclusion, we do not find signatures that suggest an alternate field-theoretic description for the SF with spatial oscillations linked to a ``symmetry enriched quantum criticality''~\cite{verresen2019gapless,thorngren2020intrinsically}.
The most striking evidence for the absence of (quasi) zero energy edge modes is provided by the finite-size scaling of the energy level splitting $\propto(L+1)^{-1}$. Due to the bulk-boundary correspondence and as outlined in \cite{thorngren2020intrinsically}, an intrinsically gapless topological phase would host edge excitations that provide strong corrections to the lowest splitting, i.e. \cref{eq:Delta E}, which we do not observe throughout the superfluid phase.
Furthermore, we do not see any spontaneous boundary occupation, nor did we observe non-vanishing edge-to-edge correlations in the thermodynamic limit.
Instead, we demonstrated the applicability of the standard Luttinger liquid description by numerical estimates of the excitation spectrum, the central charge, and the Luttinger liquid parameter.

\section{Conclusions}
\label{sec:conclusions}
In this work, we presented a state-of-the-art numerical and analytic study of the extended Bose Hubbard model in 1D. In particular, we have: 
\begin{itemize}
\item carried out detailed studies of the newly found phases.
\item confirmed and deepened the analysis of phase separation in this system, by looking at various quantities characterizing it. 
\item clarified that the entanglement spectrum oscillations survive in the thermodynamic limit and established further regions in the parameter space where this is the case.
\item not observed any edge states, or bulk-edge correspondence in these regions. Neither have we been able to propose a hidden order parameter, nor could we find a good $k$-merized variational state there, nor could we see variations from expected scaling in this universality class. 
\item confirmed that the model agrees in the superfluid phase with the predictions made within the standard framework of Luttinger-Tomonaga theory. We have provided a relation between the Luttinger parameter and the microscopic parameters of the BH model. We concluded thus the absence of topological order/effects there.

\end{itemize}

In view of recent progress with experiments on dipolar atoms, Rydberg atoms, and trapped ions, as well as novel methods of detection of entanglement entropies and spectrum, our results open an interesting playground to test CFT and Luttinger liquid properties in experiments.
Our simple bosonization approach is fully applicable in the superfluid phase only. A more powerful field theory predicting the phase transition between superfluid and supersolid would be of general interest.
The outlook for future studies includes investigations of the same model in 2D, and extension to true long-range interactions, with a particular focus on dipolar ones, where the experiments are on the way.

\begin{acknowledgements}
We like to thank P. Massignan for helpful insights and discussions. K.K. wants to thank Johannes Hauschild for his continuous help and dedication on the \textsf{TeNPy} forum.

This work was supported by the European Union’s Horizon 2020 research and innovation programme under the Marie Sklodowska-Curie grant agreement  No 713729  (K.K.), the ERC AdG's NOQIA and CERQUTE, Spanish MINECO (FIDEUA PID2019-106901GB-I00/10.13039 / 501100011033, FIS2020-TRANQI, Severo Ochoa CEX2019-000910-S and Retos Quspin), the Generalitat de Catalunya (CERCA Program,  SGR 1341, SGR 1381 and QuantumCAT), Fundacio Privada Cellex and Fundacio Mir-Puig, MINECO-EU QUANTERA MAQS (funded by State Research Agency (AEI) PCI2019-111828-2 / 10.13039/501100011033), EU Horizon 2020 FET-OPEN OPTOLogic (Grant No 899794), and the National Science Centre, Poland-Symfonia Grant No. 2016/20/W/ST4/00314, Marie Sklodowska-Curie grant STRETCH No 101029393.
\end{acknowledgements}

\bibliography{lit}

\clearpage

\newpage

\begin{appendix}

\section{Phase Separation checks}
\label{appendix:PS-size-scaling}
\begin{figure}
\centering
\includegraphics[width=.49\textwidth]{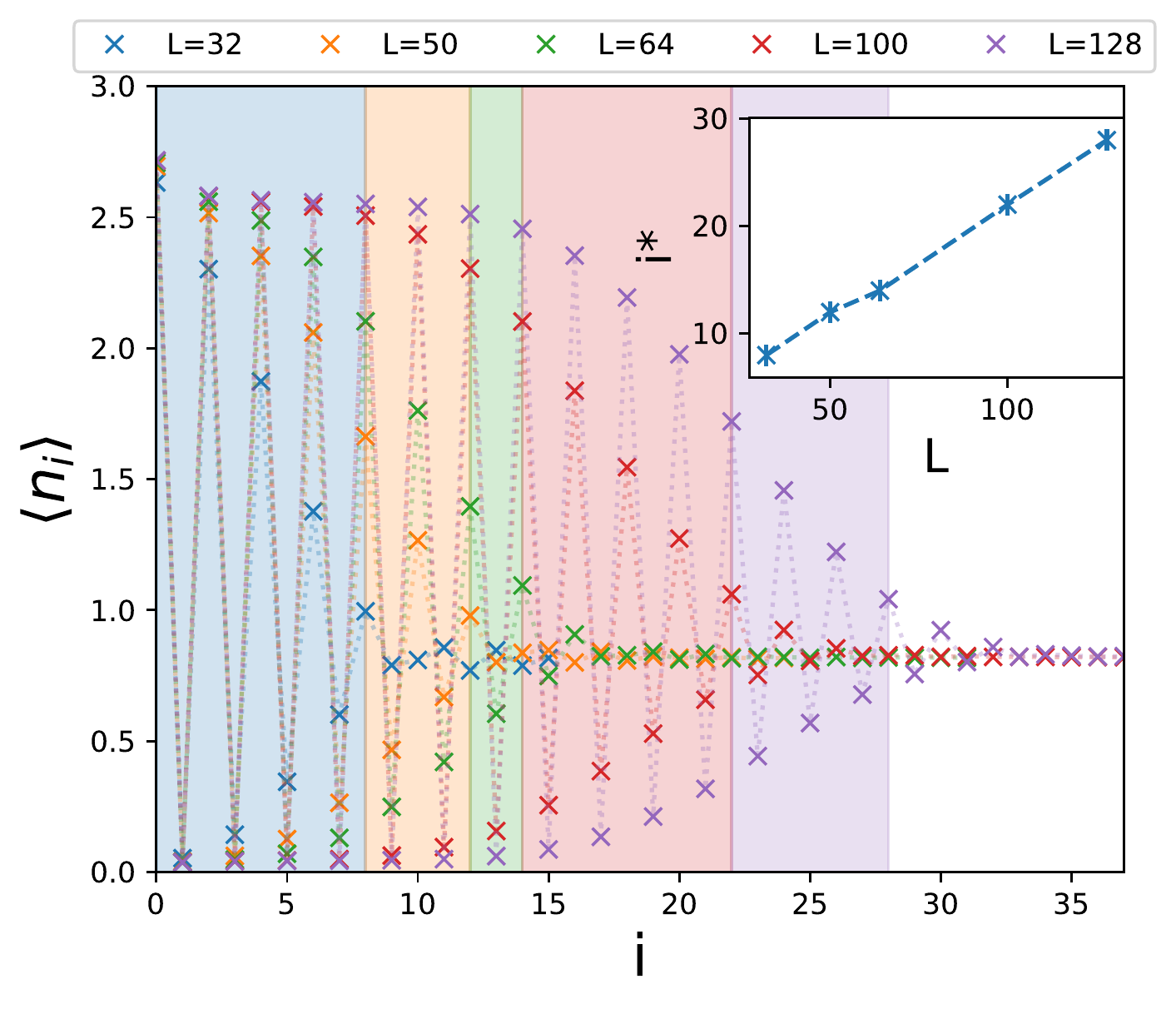}
\caption{
Linear dependence of the extent of the supersolid part in the phase-separated phase in system size. The physical parameters $(U,V,n) = (0.5,4,1)$ are fixed and the bond dimension increases with the system size with $\chimax \in [400,1000]$. The extent of the SS part $i^*$ is indicated by the colored region and estimated by calculating the maximum of the second derivative of $\braket{n_i}$ with $i$ even. The inset shows $i^*$ vs system size $L$ showing a linear dependence as expected.
}
\label{fig:appendix-PS-size-scaling}
\end{figure}

One important check for the phase separation phase is to look at how the size of the respective phases scales with $L$. In \cref{fig:appendix-PS-size-scaling} we show the extent of the supersolid phase on the boundary for different system sizes and find a linear dependence, as expected. I.e., this rules out the supersolid part being a boundary effect.

\section{Local Hilbert space dimension}
\label{appendix:d-scaling}
\begin{figure}
\centering
\includegraphics[width=.49\textwidth]{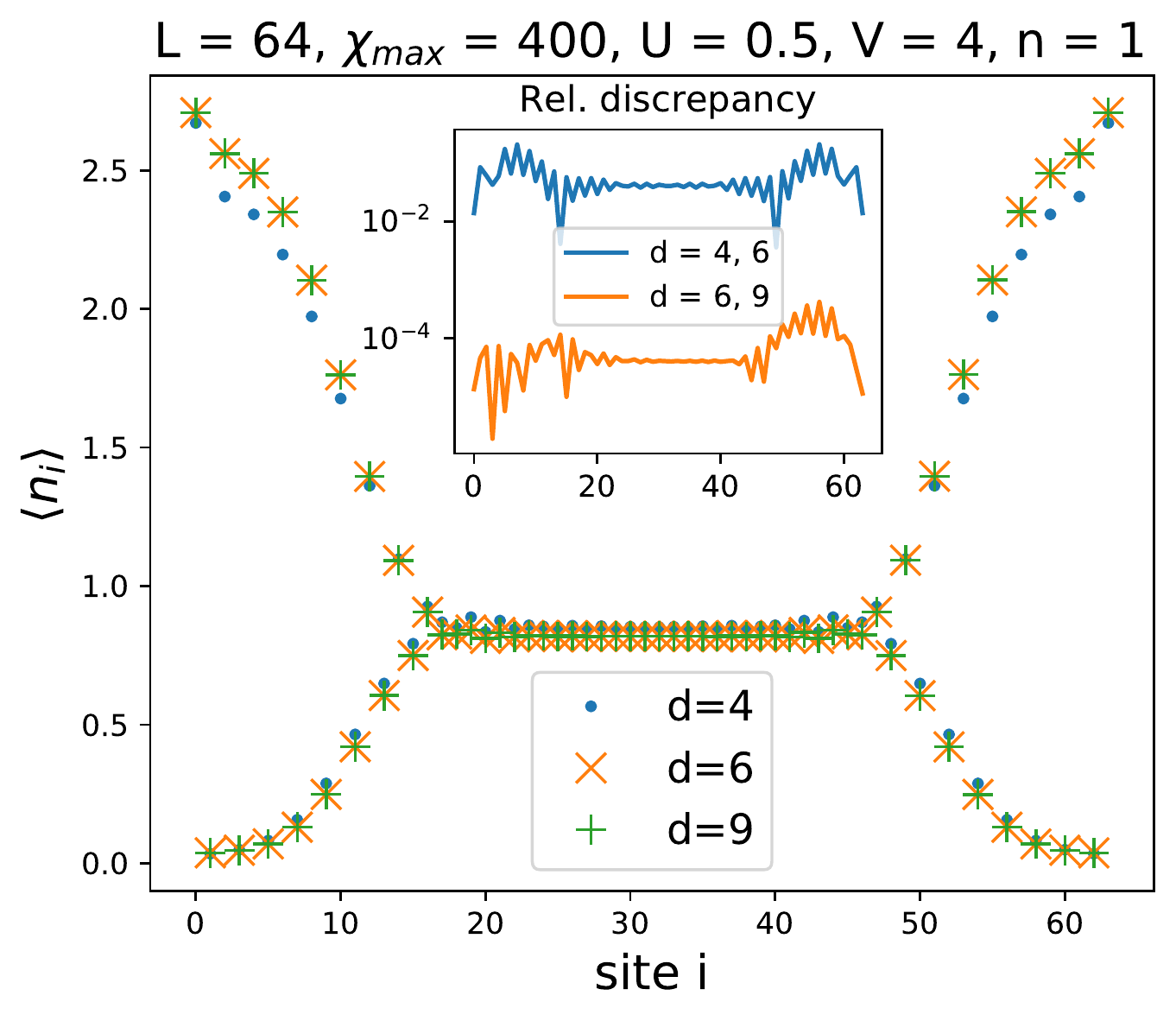}
\caption{
Comparison of the density distribution for different local Hilbert space truncations. We find no qualitative difference between $d=6$ and $d=9$. Inset: Quantitative analysis by comparing the relative discrepancy $|(\braket{n_i}^{d_1} - \braket{n_i}^{d_2})/\braket{n_i}^{d_2}|$. The absolute value of the peak deviation for $d=6,9$ is approx. $0.00025$, corresponding to a peak discrepancy of $0.04$\%.
}
\label{fig:appendix-d-scaling}
\end{figure}

\begin{figure}
\centering
\includegraphics[width=.49\textwidth]{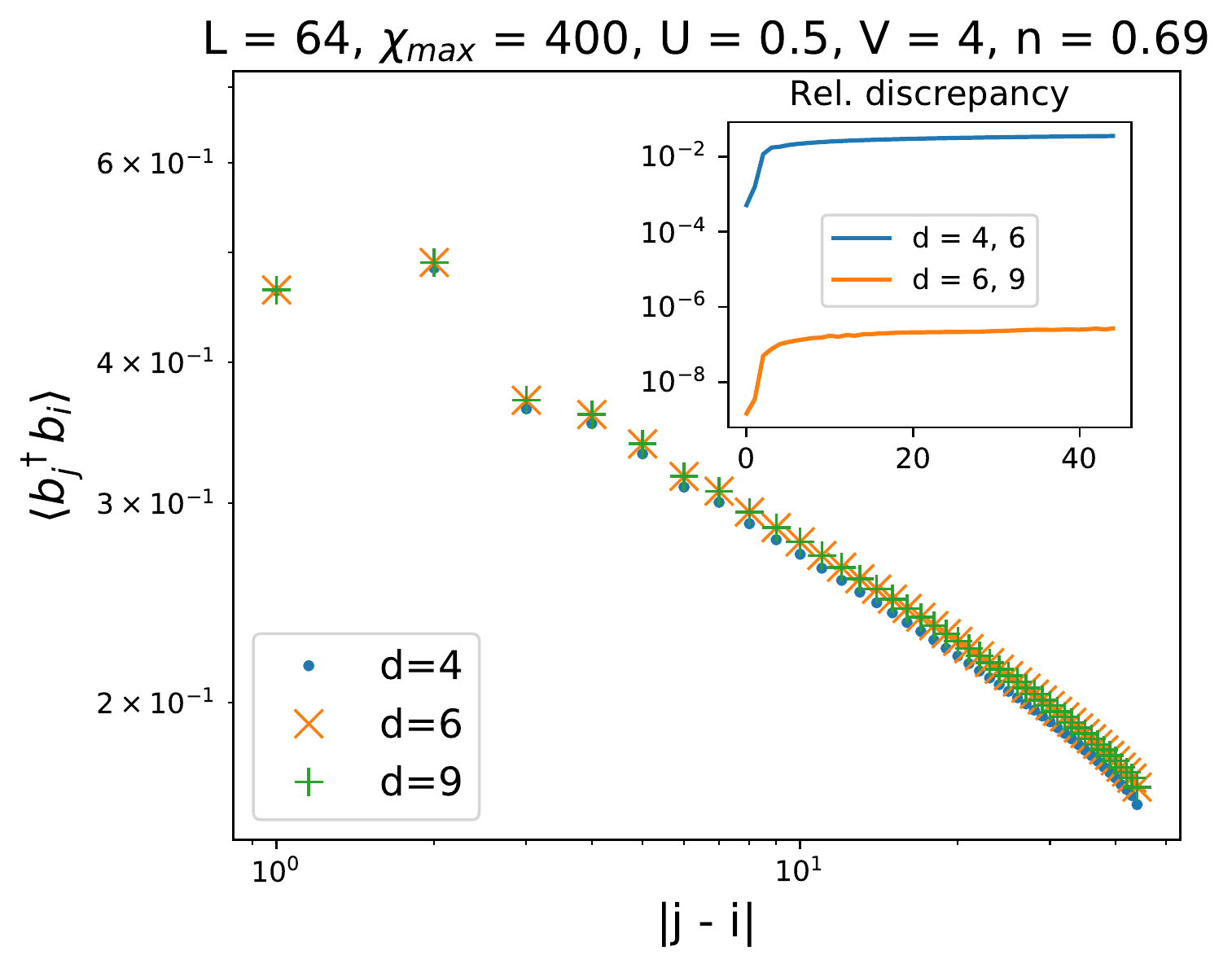}
\caption{
Comparison of $\braket{b_i^\dagger b_j}$ in the bulk for different local Hilbert space truncations. The physical parameters are set for the superfluid phase. Again, we find no qualitative difference between $d=6$ and $d=9$. Inset: Quantitative analysis by comparing the relative discrepancy.
}
\label{fig:appendix-d-scaling2}
\end{figure}

We give evidence to our claim in the main text, that we find no sgnificant difference between simulations with local Hilbert space dimension $d=6$ and $d=9$. As a first relevant example, we compare the density distribution for a fixed filling $n=1$ in the phase separated phase in \cref{fig:appendix-d-scaling}. Qualitatively, the cases $d=6$ and $d=9$ yield no visible deviation. We confirm this also quantitatively in the inset, where we see relative deviations on the order of $10^{-4}$. As a second example, we compare the correlation function $(\CSF)_{ij} = \braket{b_j^\dagger b_i}$ for a superfluid ground state with again the same local Hilbert space dimensions $d$ in \cref{fig:appendix-d-scaling2}. We find again no qualitative differences between $d=6$ and $d=9$, as well es quantitative discrepancies on the order of $10^{-8}$. Overall, we conclude that $d=6$ is sufficient for the effects that we study in the main text.

\section{Comparison SF and SS}
\begin{figure}
\centering
\includegraphics[width=.49\textwidth]{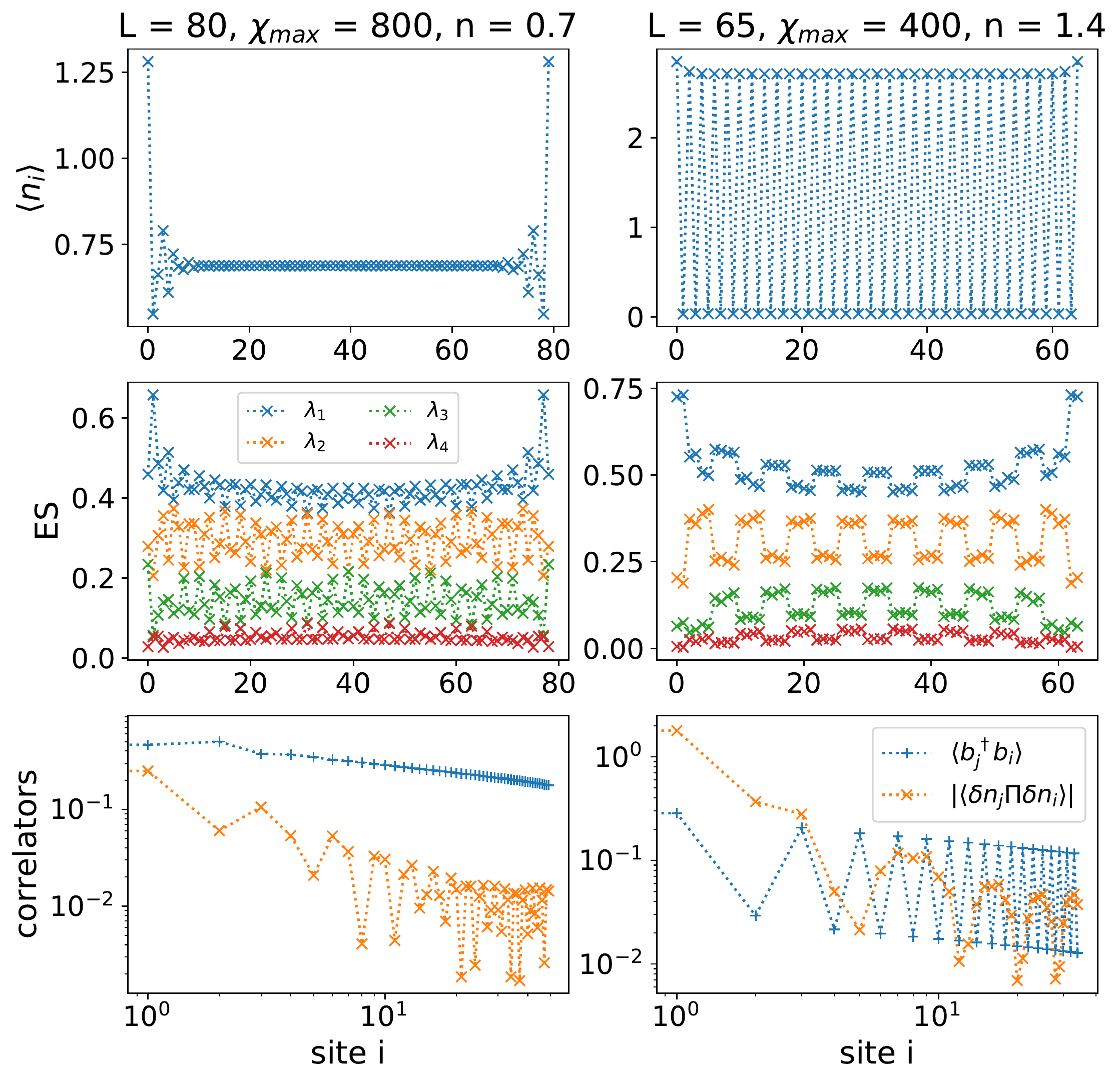}
\caption{
Comparison of the SF (left panel) and SS (right panel) phase at incommensurate fillings with oscillating patterns. a) The main difference of both phases is the flat vs. the solid pattern in the spatial density. b) The entanglement spectrum (ES) shows spatial oscillations. In the case of the SS, the oscillating nodes are pairs of sites. c) Both phases are superfluid with an algebraic decay of $\CSF = \braket{b_j^\dagger b_i}$. Further, the hidden spatial oscillations can be unveiled by measuring $\Cst = \braket{\delta n_j \Pi \delta n_i}$ with $\Pi = \exp\left(-i \pi \sum_{0 \leq l < j} \delta n_l\right)$ and $\delta n_l =  n_l - n$. Note that the frequency and shape of these oscillations changes with the filling $n$, what we display are just two examples.
}
\label{fig:appendix-SF-SS}
\end{figure}

We give more details on the characterization of the (homogeneous) superfluid and supersolid phases at incommensurate fillings. 
The main property that distinguishes the superfluid for fillings below the phase separation phase and supersolid for fillings above the phase separation phase, is the solid pattern in density as can be seen in \cref{fig:appendix-SF-SS}. However, they share unexpected properties that manifest themselves in the entanglement spectrum or string correlators as is discussed in the main text for the SF phase. The main property is that the entanglement spectrum shows spatial oscillations. In the case of the SS, the oscillating nodes are pairs of two sites. Further, both phases are superfluid with an algebraic deca in $\CSF = \braket{b_j^\dagger b_i}$. In both cases, the hidden pattern can be unveiled by looking at the string order correlator $\Cst = \braket{\delta n_j \Pi \delta n_i}$ with $\Pi = \exp\left(-i \pi \sum_{0 \leq l < j} \delta n_l\right)$ and $\delta n_l =  n_l - n$.

\section{Translational invariance of the superfluid phase}
\begin{figure*}
\centering
\includegraphics[width=.85\textwidth]{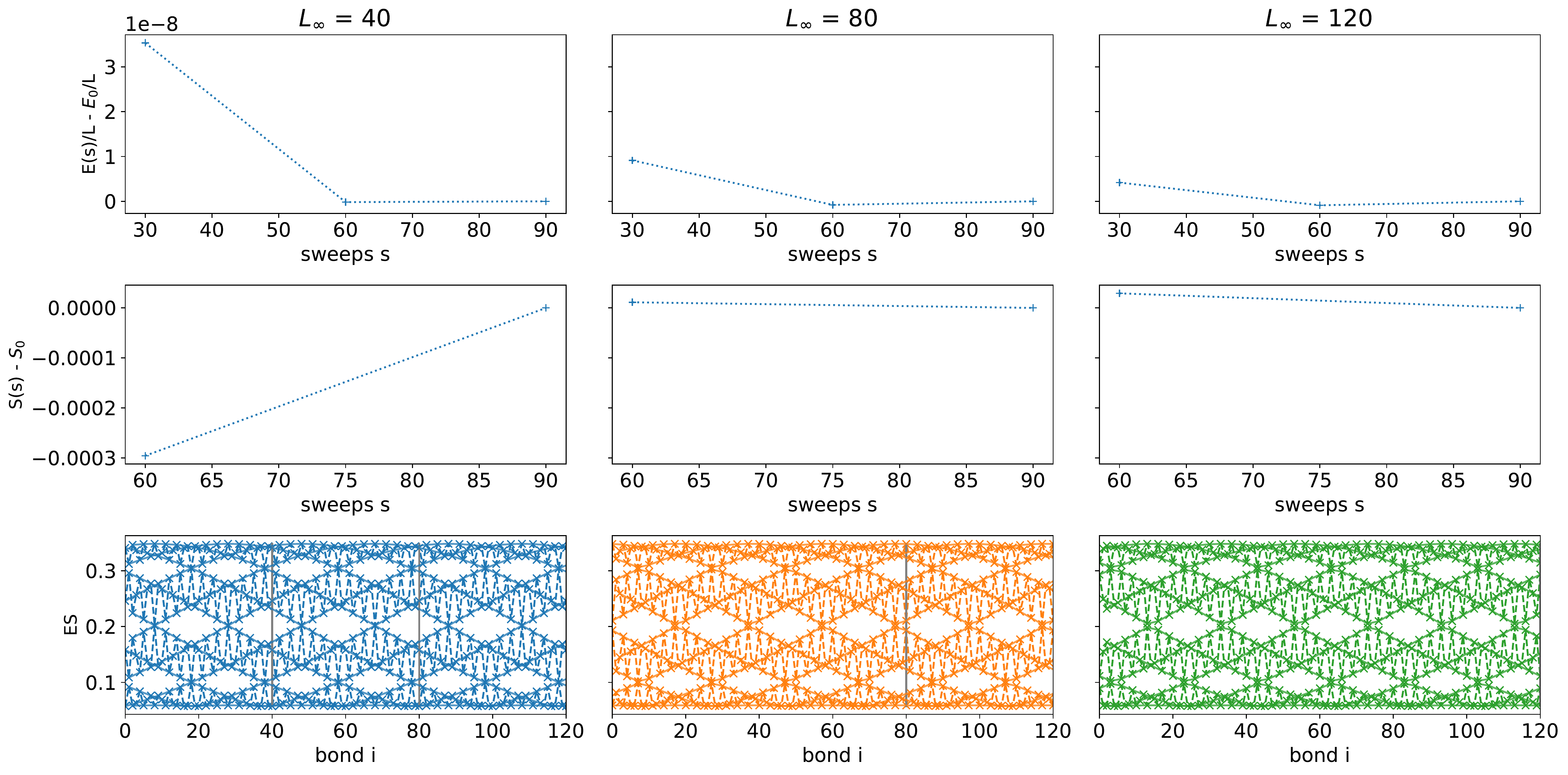}
\caption{
iDMRG convergence for the filling $n=13/20=0.65$. We check the convergence of the energy per site $E/L_\infty$ and mean entanglement entropy $S$ with respect to the number of sweeps through the unit cell for sizes $L_\infty = 40, 80, 120$. The bottom row shows the emerging entanglement spectrum that matches perfectly in all three cases. The used bond dimension is $\chimax = 400$ and the convergence criteria are to either have the relative change in energy be smaller than $10^{-10}$ or reaching a maximum number of $1000$ sweeps. The vertical grey line indicates where the unit cell is repeated.
}
\label{fig:appendix-iDMRG-0.65}
\end{figure*}

\begin{figure*}
\centering
\includegraphics[width=.85\textwidth]{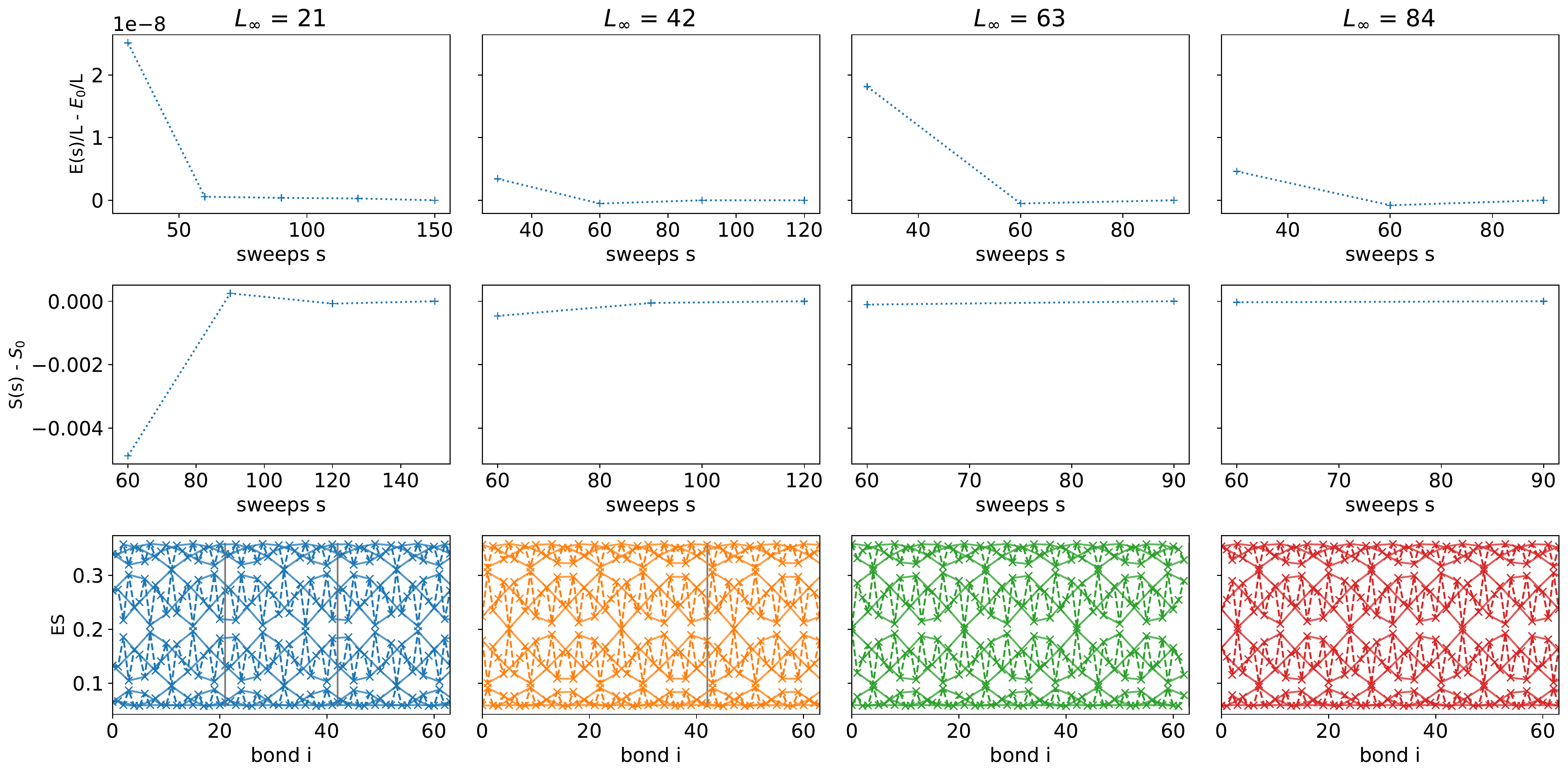}
\caption{
iDMRG convergence for the filling $n=13/21\approx 0.619$. We check the convergence of the energy per site $E/L_\infty$ and mean entanglement entropy $S$ with respect to the number of sweeps through the unit cell for sizes $L_\infty = 21, 42, 63, 84$. The bottom row shows the emerging entanglement spectra that match well in all four cases. The used bond dimension is $\chimax = 400$ and the convergence criteria are to either have the relative change in energy be smaller than $10^{-10}$ or reaching a maximum number of $1000$ sweeps.
}
\label{fig:appendix-iDMRG-0.62}
\end{figure*}

\begin{figure*}
\centering
\includegraphics[width=.7\textwidth]{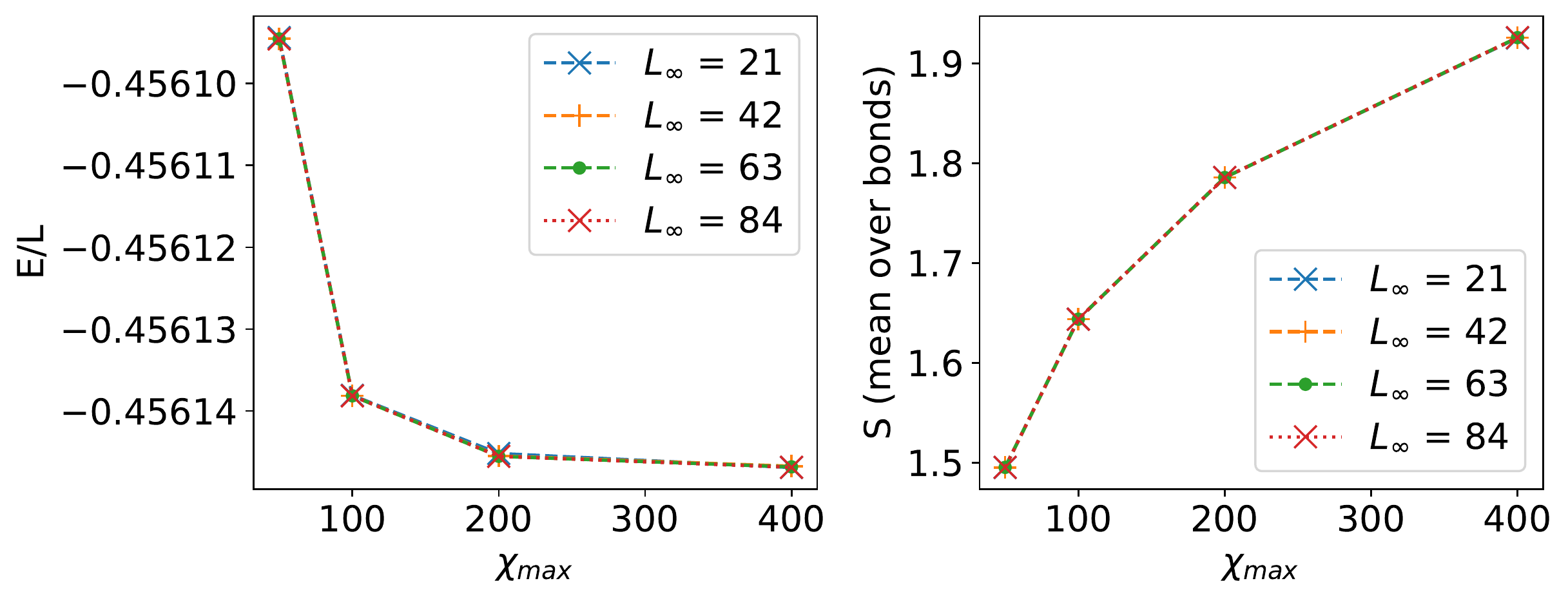}
\caption{
Energy per site and mean entanglement entropy $S$ for different bond dimensions $\chimax$. In terms of energy, we are reaching saturation for the bond dimension $\chimax = 400$ that we are using. In terms of the entanglement entropy, we see a logarithmic grows as is expected for critical phases.
}
\label{fig:appendix-iDMRG-0.62_bonds}
\end{figure*}

\begin{figure*}
\centering
\includegraphics[width=.85\textwidth]{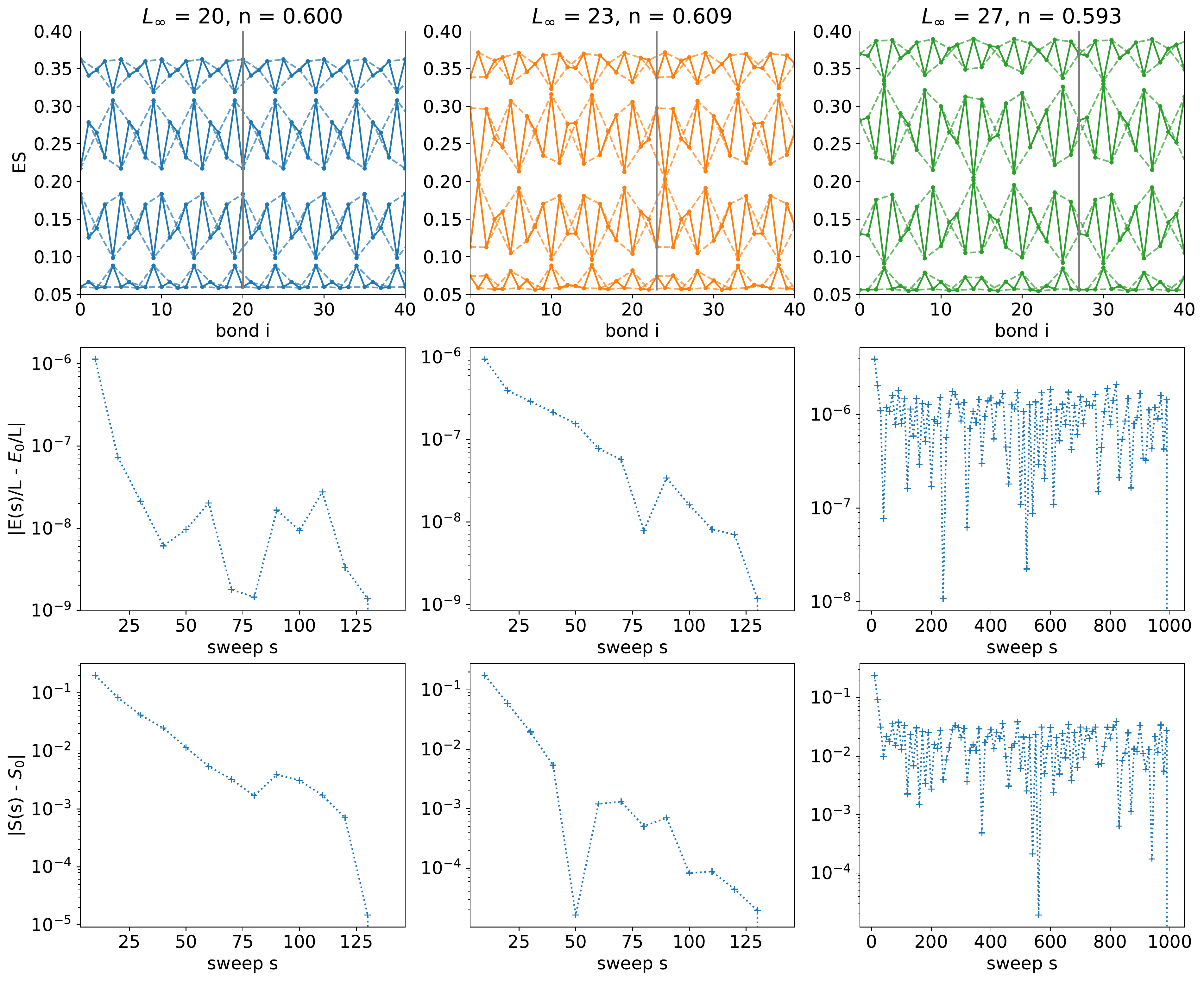}
\caption{
iDMRG convergence for the filling $n=3/5 = 0.6$. The convergence criteria of relative change in energy be smaller than $10^{-10}$ is not met for $L_\infty = 27$ with the optimization terminating at the maximum number of sweeps $1000$. For $L_\infty=20, 23$ the criteria is met, however the optimization is not monotonic. From the entanglement spectra, it seems there is some strain and the unit cell sizes cannot accommodate the desired spatial period the system is trying to establish.
}
\label{fig:appendix-iDMRG-0.6}
\end{figure*}

\begin{figure*}
\centering
\includegraphics[width=.99\textwidth]{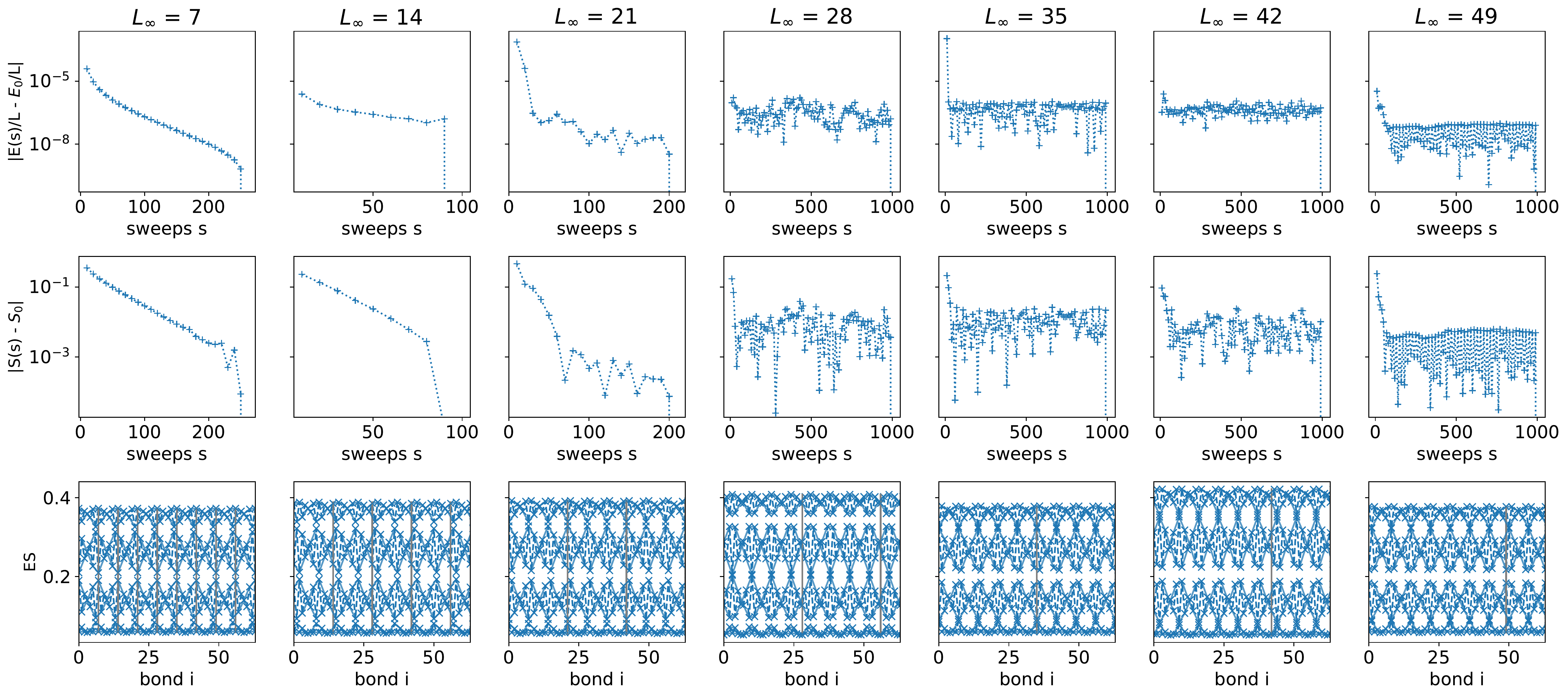}
\caption{
iDMRG convergence for the filling $n=4/7\approx0.571$ for system sizes that are multiples of $7$. Again, we find that strain is causing convergence problems. For $L_\infty = 7, 14, 21$ it meets the convergence criteria but fails for $L_\infty = 28$ and beyond. The convergence for $L_\infty = 7, 14, 21$ might be misleading and might only arise because of a lack of space for strain to emerge.
}
\label{fig:appendix-iDMRG-0.571}
\end{figure*}

\begin{figure*}
\centering
\includegraphics[width=.7\textwidth]{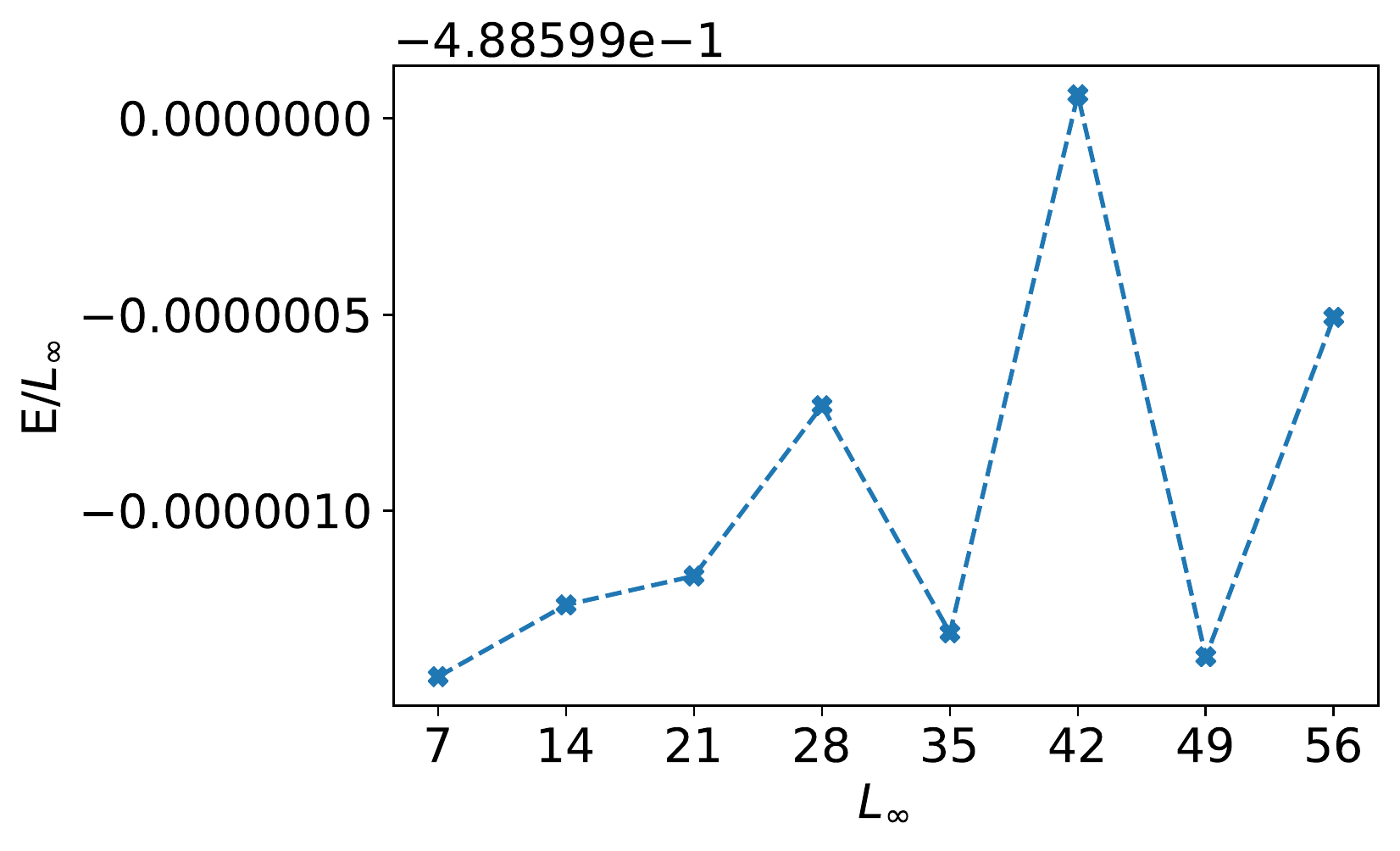}
\caption{
Energy per site for $n=4/7$ with the unit cell being a multiple of $7$. Corresponding convergences are displayed in \cref{fig:appendix-iDMRG-0.571}. Despite strain and convergence problems, the finale ground state energies are still well within range of each other.
}
\label{fig:appendix-iDMRG-0.571-energy}
\end{figure*}

\begin{figure*}
\centering
\includegraphics[width=.85\textwidth]{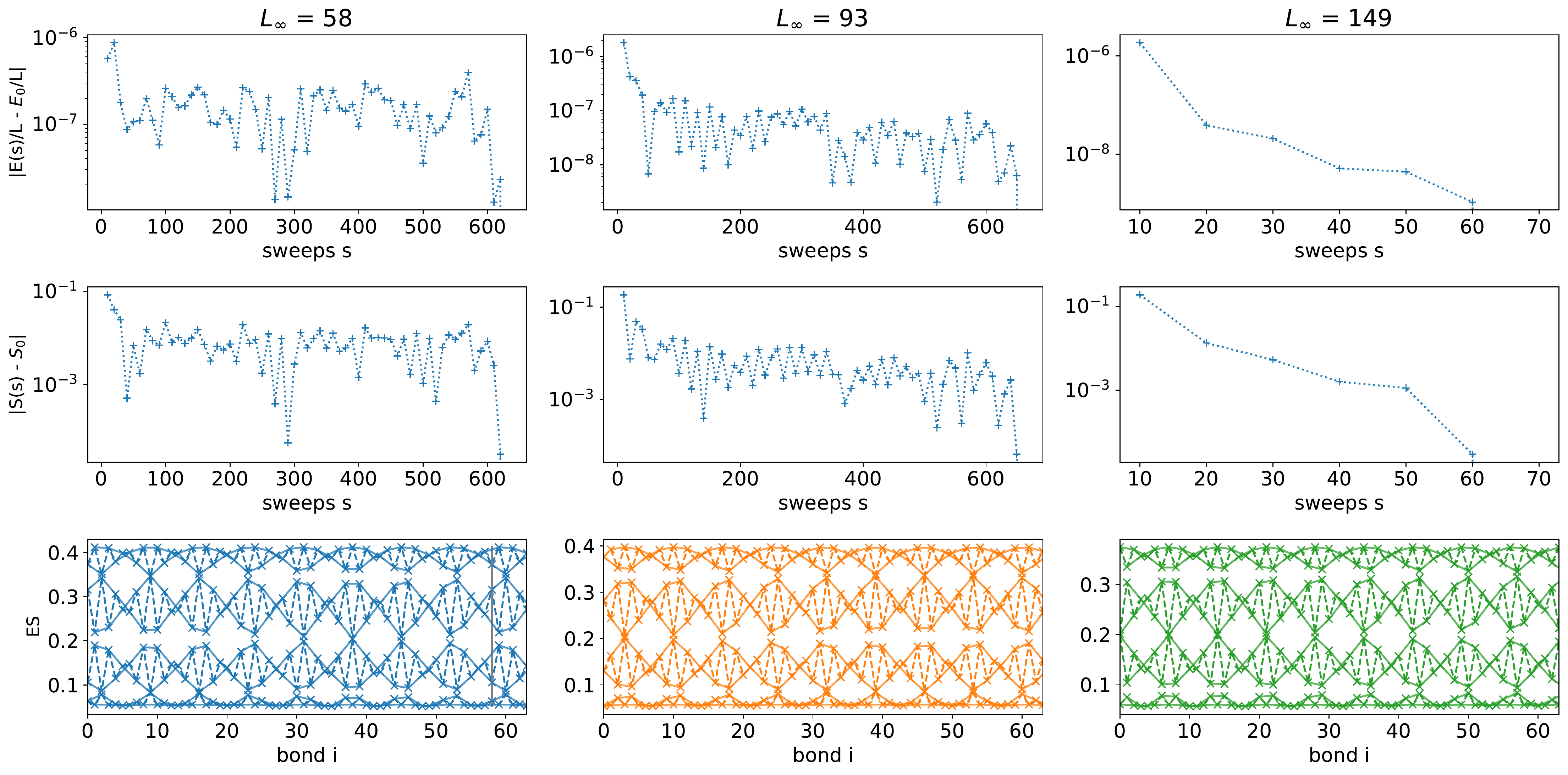}
\caption{
iDMRG convergence for $n=\approx4/7\approx0.571$ with system sizes that are not multiples of $7$. We look at $n=33/58\approx0.569$, $n=53/93\approx0.570$ and $n=85/149\approx0.570$. In all cases the convergence criteria is eventually met. However, the irregular entanglement spectra patterns indicate that also here the unit cell sizes cannot accommodate well the desired spatial pattern.
}
\label{fig:appendix-iDMRG-0.571-incommensurate}
\end{figure*}

\begin{figure*}
\centering
\includegraphics[width=.85\textwidth]{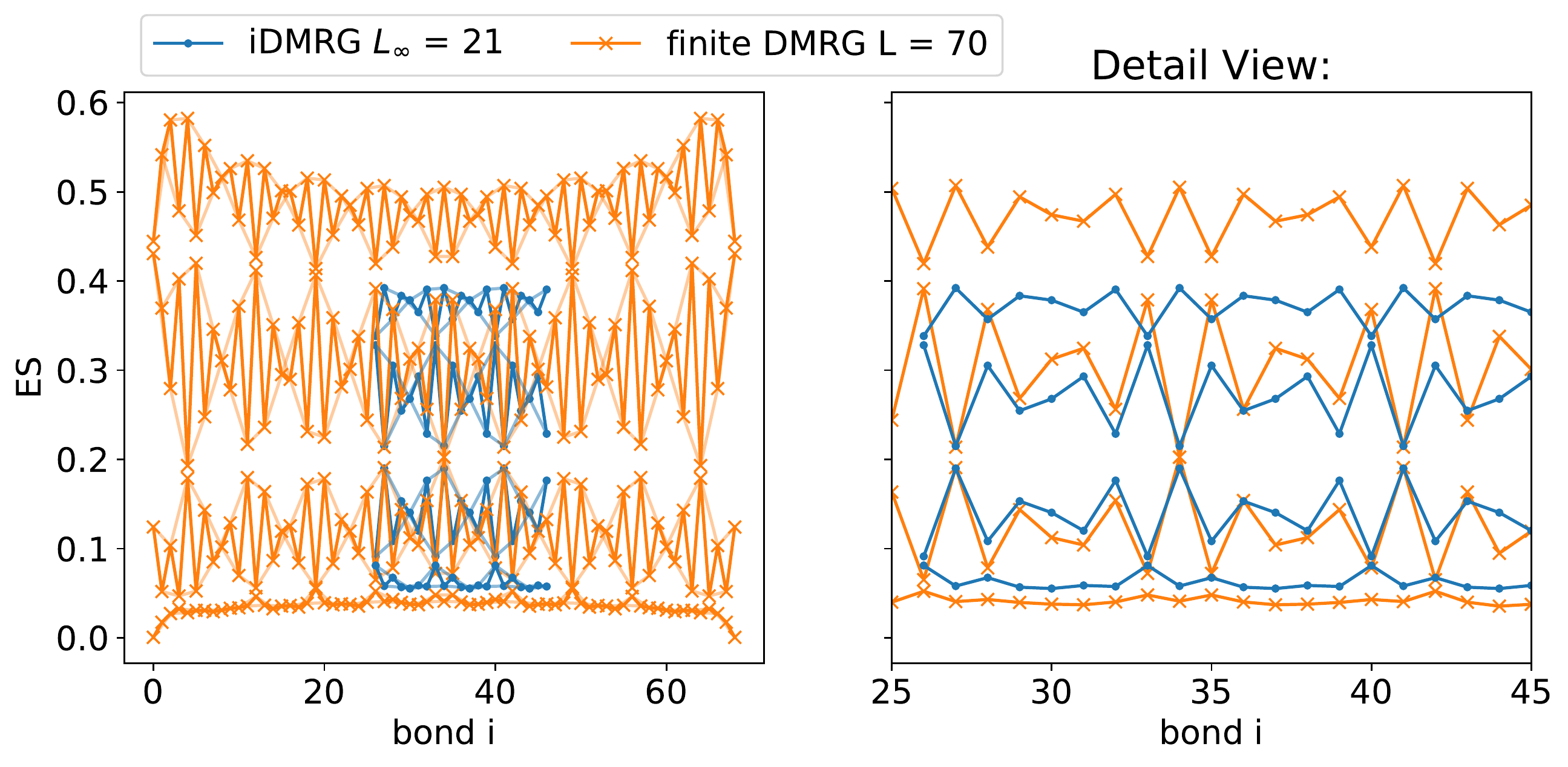}
\caption{
Comparison of finite and infinite DMRG simulations for $n=4/7$ with strain. Even though the unit cell size is not optimal for the targeted filling, the spatial pattern of the infinite system is still matching well with the bulk of the finite system.
}
\label{fig:appendix-iDMRG-0.571-comparison}
\end{figure*}

The emergence of the spatial pattern in the superfluid and supersolid phase close to the phase separation is a manifestation of a broken translational symmetry. To further strengthen this point, we perform some checks with iDMRG. We expect that iDMRG is running into problems when the chosen unit cell size is incommensurate with the spatial period of the entanglement spectrum of the system. This is indeed what we find. The spatial period depends on the targeted filling. For some fillings, and thus for some spatial periods, it has proven harder or easier to find a suitable unit cell size.
We start with the case $n=13/20=0.65$ that is shown in the main text. We tried with $L_\infty = 40, 80, 120$ and reached good convergence in all cases as is displayed in \cref{fig:appendix-iDMRG-0.65}. There is no sign of strain as the emerging pattern is regular and repeats perfectly. In this case, the bond dimension is chosen to be. Similarily, we get good results for $n = 13/21 \approx 0.619$ as shown in \cref{fig:appendix-iDMRG-0.62} for $L_\infty = 21, 42, 63, 84$. We also check the dependence on the bond dimension in \cref{fig:appendix-iDMRG-0.62_bonds} and find sufficient convergence for $\chimax = 400$, that we use throughout this analysis.
We can see how strain is manifested in the ES pattern and how this leads to convergence problems in the example of $n=3/5=0.6$ in \cref{fig:appendix-iDMRG-0.6}. For $L_\infty = 27$ it does not converge at all, the energy is oscillating until the maximum number of sweeps (1000) is reached. For $L_\infty=20$ and $23$, the convergence criteria is met, but we can still see some non-monotonicities - however on a much smaller scale. It seems there is a certain tolerance to strain when the missmatch is not too large.
For filling $n=4/7=0.579$ it proved very difficult to find a suitable unit cell size. We check for unit cells that are multiples of $7$ in \cref{fig:appendix-iDMRG-0.571}. While the convergence criteria is not met for $L_\infty = 28$ and beyond, we still find comparable ground state energies for uneven numbers of sites, as seen in \cref{fig:appendix-iDMRG-0.571-energy}. We also tried for system sizes that are not multiples of $7$ in \cref{fig:appendix-iDMRG-0.571-incommensurate}. While the convergence criteria is eventually met here, the irregular pattern in the entanglement spectrum indicates that also here the unit cell sizes are not optimal to accommodate the desired spatial pattern. We conjecture that with increasing unit cell sizes, a suitable size can be approached, but is in practice difficult or infeasible to simulate due to large system sizes. Curiously, even for these incommensurate sizes, the match between infinite and finite DMRG is still well, as can be seen in \cref{fig:appendix-iDMRG-0.571-comparison}.

To summarize: When simulating the system with iDMRG, one encounters problems with convergence whenever the unit cell size is incommensurate with the spatial period of the system, and converges well when the unit cell size is commensurate. Overall, this behavior is what one would expect of a system with broken translational symmetry strengthens the point that the effect is indeed physical.

\end{appendix}~
\end{document}